\documentclass[final,3p,11pt,times,twocolumn]{elsarticle}

\usepackage{amsmath,amssymb}
\usepackage[table,xcdraw]{xcolor}
\usepackage{booktabs,multirow}
\usepackage{subcaption}
\usepackage{lineno}
\usepackage{comment}

\makeatletter
\def\ps@pprintTitle{%
   \def\@oddhead{}%
   \let\@evenhead\@oddhead
   \def\@oddfoot{}%
   \let\@evenfoot\@oddfoot
}
\makeatother

\usepackage{subcaption}

\begin{document}

\begin{frontmatter}



\title{Nested Control Co-design of a Spar Buoy Horizontal-axis Floating Offshore Wind Turbine}


\author[author-ise]{Saeid Bayat\corref{author-corr}}
\author[author-uofm]{Yong Hoon Lee}
\author[author-ise]{James T. Allison}

\affiliation[author-ise]{
    department = {Industrial and Enterprise Systems Engineering,},
    organization = {University of Illinois at Urbana-Champaign},
    city={Urbana},
    state={IL},
    country={USA}
}
\affiliation[author-uofm]{
    department = {Mechanical Engineering,},
    organization = {The University of Memphis},
    city={Memphis},
    state={TN},
    country={USA}
}

\cortext[author-corr]{Corresponding author. Email: Bayat2@illinois.edu, Address: 509 E White, Unit 1, Chmpaign, IL 61820}

\begin{abstract}
Floating offshore wind turbine (FOWT) systems involve several coupled physical analysis disciplines, including aeroelasticity, multi-body structural dynamics, hydrodynamics, and controls. Conventionally, physical structure (plant) and control design decisions are treated as two separate problems, and generally, control design is performed after the plant design is complete. However, this sequential design approach cannot fully capitalize upon the synergy between plant and control design decisions. These conventional design practices produce suboptimal designs, especially in cases with strong coupling between plant and control design decisions. Control co-design (CCD) is a holistic design approach that accounts fully for plant-control design coupling by optimizing these decisions simultaneously. CCD is especially advantageous for system design problems with complex interactions between physics disciplines, which is the case for FOWT systems. This paper presents and demonstrates a nested CCD approach using open-loop optimal control (OLOC) for a simplified reduced-order model that simulates FOWT dynamic behavior. This simplified model is helpful for optimization studies due to its computational efficiency, but is still sufficiently rich enough to capture important multidisciplinary physics couplings and plant-control design coupling associated with a horizontal-axis FOWT system with a spar buoy floating platform. The CCD result shows an improvement in the objective function, annual energy production (AEP), compared to the baseline design by more than eleven percent. Optimization studies at this fidelity level can provide system design engineers with insights into design directions that leverage design coupling to improve performance. These studies also provide a template for future more detailed turbine CCD optimization studies that utilize higher fidelity models and design representations.

\end{abstract}


\end{frontmatter}

\section{Introduction}
\label{sec:intro}

Over the last decade, offshore wind energy has gained significant attention as an environmentally friendly and cost-effective energy source. Offshore wind resources are stronger and more consistent than their onshore counterparts, while exhibiting fewer issues, such as terrain-induced shear effects, acoustic noise, visual impact, size limits, and interference with humans and wildlife. However, many ideal offshore wind locations have water depths beyond 45 meters, rendering monopile foundations installed directly on the sea bed unsuitable \citep{Bhattacharya2014Foundation}. Therefore, there is a growing need for technical advancements in floating offshore wind turbine (FOWT) systems to harness the profound energy potential in deep water regions.

FOWTs are still in their early stages of development and have yet to be commercially deployed. To facilitate large-scale commercialization of the FOWTs, several challenges must be addressed, which can be mitigated through innovative design discoveries and optimization strategies that leverage inherent complex design couplings and significantly reduce energy costs. For example, advanced control strategies can help enhance energy production while simultaneously protecting the turbine system from motions induced by hydrodynamic interactions with the floating platform. Furthermore, FOWT systems consist of numerous components with multidisciplinary interactions, including hydrostatics, hydrodynamics, mooring dynamics, structural dynamics, aerodynamics, and controls, complicating the design process. Thus, advancing FOWT performance requires advanced multidisciplinary design methodologies.

This article presents a study of the integrated physical (plant) and control design of a spar buoy-based horizontal-axis FOWT system using a model comprised of a set of simplified governing physics formulations. The FOWT model development is intended for use with integrated physical and control design optimization (often referred to as a control co-design, CCD), which utilizes coupling between plant and control design variables to better exploit the properties of the combined plant-control design space. 
The design problem is simple enough to directly demonstrate how FOWT CCD can be formulated but contains enough richness to explore important physics and design couplings.

Direct optimal control methods have proven particularly effective in the flexible exploration of new control strategies while accounting for the realities of physical system design, such as design coupling and failure modes \cite{Allison2014JMDCCD, Allison2014a, Bohme2017IDM, Herber2019CCD,bayat2023ss,SaeidBayat-Vehicle}.
Direct transcription is a class of direct optimal control methods that overcomes many of the shortcomings of conventional shooting-based direct optimal control methods \citep{Betts1998Survey} but requires that optimization solvers have direct access to the time derivative function. Instead of treating system dynamics as a black box, direct transcription capitalizes on the problem structure inherent to continuous dynamic system optimization problems to enhance numerical robustness and efficiency.
Here, in this study, we use pseudospectral-based direct transcription method \citep{Ross2012PSreview, Patterson2014GPOPS2, Garg2009PS3} for obtaining optimal control solutions.

Engineered systems are often operated using active control systems, and designing these systems requires engineers to make both physical system (plant) and control system design decisions.
Plant-control design coupling exists if changing plant design influences what control decisions are best for overall system performance, and if control design changes influence what plant design decisions are best. Despite the coupling mentioned above, a sequential procedure has largely been used to design these systems in practice (control design performed after completing plant design). Sequential approaches, however, leave potential performance gains on the table by not fully exploiting design coupling to produce system-optimal designs \citep{Fathy2001CCD}.

Here, an integrated design approach is used that fully accounts for bi-directional plant-control design coupling and produces system-optimal designs, generally referred to as contro co-design (CCD). The CCD optimization is a class of integrated design methods that account explicitly for plant-control coupling to support the discovery of non-obvious system designs that realize new performance levels \citep{Allison2014JMDCCD, Herber2019CCD, Fathy2001CCD, Peters2009CCD, Allison2010CCD, Lee2019WCSMO}. Design coupling is accounted for when plant and control design decisions are considered simultaneously in the system optimization problem. Variations on the simultaneous CCD formulation, such as nested or distributed optimization, have been studied that provide numerical advantages over the simultaneous approach while being mathematically equivalent \citep{Allison2014JMDCCD, Herber2019CCD, Allison2010CCD, Lee2019WCSMO, Reyer2000CCD}.

The importance of CCD in wind energy research has gained recent attention \citep{garcia2019control}. A holistic consideration of control and physical systems in a CCD framework provides significant possibilities of effective and cost-efficient wind turbine system design discoveries. The potential performance gains are more pronounced in FOWTs than in land-based turbines due to more complex couplings that arise when wind turbines are installed on floating platforms.
Various multidisciplinary design optimization (MDO) architectures, including monolithic methods \citep{ashuri2014multidisciplinary, du2020control} or nested methods \citep{deshmukh2016multidisciplinary} are implemented to explore potential design advancements in wind energy systems. The results of these studies demonstrated that the consideration for design coupling provides the large advantages over designing for each disciplinary domain sequentially.

While previous studies have explored CCD for wind turbines, they have yet to investigate the potential value of applying the CCD to FOWTs. Fairly comprehensive CCD studies have been performed specifically for land-based wind turbines but often relied on computationally-expensive models. Due to the increased computational expenses beyond land-based models, many previous CCD studies on FOWTs are limited to non-holistic approaches. While a handful of articles present integrated design approaches for FOWTs, most utilize sequential design approaches with open-loop optimal control (OLOC) or model predictive control (MPC). This study fills a critical gap in integrated FOWT system design and provides an initial foundation for future studies in this area.

In this study, we demonstrate a CCD FOWT implementation that utilizes a reduced-order model (ROM) that emulates the dynamic behaviors of a horizontal-axis FOWT installed on a spar buoy floating platform. The baseline design adapts specifications of the NREL 5 MW reference wind turbine \citep{Jonkman2009NREL5MW} installed on the OC3-Hywind spar buoy platform \citep{Jonkman2010OC3Hywind}. We use a nested CCD approach that accounts for plant-control design coupling while capitalizing on specialized solution strategies for direct optimal control. The outer-loop optimization problem navigates the plant design space using the covariant matrix adaptation evolution strategy (CMA-ES), which is a gradient-free population-based optimization algorithm that utilizes a covariant matrix. The covariant matrix adaptation procedure utilizes an estimation of the inverse Hessian for a quadratic function to further evolve sampling shape toward the descent direction \citep{Hansen2003CMAES, Hansen2016CMAES}. The inner-loop OLOC problem finds the optimal state and control trajectories for a candidate plant design specified by the outer loop. The OLOC problem is discretized using the pseudospectral direct optimal control solver, GPOPS-II \citep{Patterson2014GPOPS2}, and numerically solved using the interior-point method NLP solver IPOPT \citep{Wachter2006IPOPT}.

Our unique contributions presented in this study include (1) a demonstration of the holistic nested CCD design procedure for the FOWT system plant and control design variables, (2) a computationally efficient FOWT model based on reduced states and degrees-of-freedom (DOF) for system dynamics, including neural network-based mooring line dynamics, (3) implementation of CMA-ES, a gradient-free method, in the outer-loop plant design optimization to tackle rugged numerical responses induced by the inner-loop OLOC solution, and (4) analyses of the design solution that provide in-depth knowledge and insights into the coupled effects of the design parameters and the FOWT system behavior. Section~\ref{sec:prob-def} presents the problem definition, including the FOWT model, design variables, and objective and constraint functions. Section~\ref{sec:results} presents the results and analysis of two CCD case studies: (1) optimization of tower and control design variables and (2) optimization of the tower, blades, and control design variables. Finally, Section \ref{sec:conclusion} presents the concluding remarks of this study.

\section{Problem Definition}
\label{sec:prob-def}

\subsection{{FOWT} Dynamic Model and Plant Design Variables}
\label{sec-sub:overall-model}

FOWTs can utilize different floating platform types, including, but not limited to, spar buoy, semi-submersible, tension leg platforms (TLPs), and barge platforms. The studies presented here is based on a spar buoy platform. Because many different disciplines and components are interacting with each other in the system, deriving detailed dynamical equations is complicated. In addition, as we increase the level of complexity, more advanced models are needed, and as a result, computational time can increase significantly. Simulating dynamic behaviors with such high complexity can be achieved using sophisticated aero-hydro-servo-elastic simulation tools, such as OpenFAST \citep{Jonkman2013,OpenFAST}. Working to solve CCD optimization problems by linking optimization solvers directly to simulation tools such as OpenFAST can enhance design problem fidelity, but presents several difficult challenges. For example, direct transcription may not be possible to use with simulation tools that can only be run as a black box, necessitating the use of the more limited single shooting method. Recent work in the use of adaptive surrogate models for time derivative functions has made possible the use of black-box simulation tools with direct transcription, but is very limited in problem dimension \cite{Des17d}. A dynamic system model is needed that provides direct access for the optimization solver to the time derivative function. This need is addressed here by constructing a reduced-order FOWT model that is compatible with CCD methods based on direct transcription, while providing sufficient richness and accuracy to support useful CCD optimization studies. 

Several previous efforts have produced reduced-order FOWT models. For example, \citet{lemmer2018low} developed a simplified model for FOWTs with semi-submersible platforms, and \citet{al2017dynamics} developed a similar dynamic model for spar buoy FOWTs. In this work, we have adapted the model published by \citet{al2017dynamics} to obtain a spar buoy FOWT model that meets CCD optimization study requirements. This implementation utilizes analytical dynamics to derive the equations of motion in terms of platform quasi-coordinates. A two-dimensional dynamic equation with an assumption of a rigid tower is introduced in Eq.~\eqref{eq:dynamics}:

\begin{align}
\label{eq:dynamics}
\begin{split}
&\begin{bmatrix}
     M_{\text{sys}}+\bar{A}
\end{bmatrix}_{4\times 4}
\begin{bmatrix}
     \dot{v}_x\\
     \dot{v}_z\\
     \dot{\omega}_y\\
     \dot{\Omega}
\end{bmatrix}_{4\times 1}
+
\begin{bmatrix}
    \Tilde{S} M_{\text{sys}}+\bar{C}_A
\end{bmatrix}_{4\times 4}
\begin{bmatrix}
     v_x\\
     v_z\\
     \omega_y\\
     \Omega
\end{bmatrix}_{4\times 1} \\
&=
\begin{bmatrix}
     F^m_{2\times 1}\\
     M^m_{1\times 1}\\
     0
\end{bmatrix}_{4\times 1}
+
\begin{bmatrix}
     F_{2 \times 1}\\
     M_{1\times 1}\\
     \tau_{\text{a}} - \tau_{\text{g}}
\end{bmatrix}_{4\times 1}
\end{split}
\end{align}

where $M_{\text{sys}}$ is the mass matrix, $\bar{A}$ is the hydrodynamic added mass matrix, $\bar{C}_A$ is the hydrodynamic Coriolis and centripetal matrix, $v_x$ and $v_z$ are the platform velocities in the surge and heave directions, respectively, $\omega_y$ is the platform pitch rate, $\Omega$ is rotor rotational speed, $F^m$ is the gravitational force in the surge and heave directions, $M^m$ is the gravitational moment in the pitch direction, $F$ quantifies the external forces in the surge and heave directions, $M$ is the external moment in the pitch direction, $\tau_{\text{a}}$ is the aerodynamic torque, and $\tau_{\text{g}}$ is the generator torque. $\Tilde{S}$ is also a matrix that is a function of $v_x$, $v_z$, and $\omega_y$, and is presented in Eq.~\eqref{eq:S-tilde}:

\begin{align} \label{eq:S-tilde}
    \Tilde{S}=
    \begin{bmatrix}
        0 & \omega_y & 0 & 0\\
        -\omega_y & 0 & 0 & 0\\
        v_z & -v_x & 0 & 0\\
        0 & 0 & 0 & 0
    \end{bmatrix}
\end{align}

$M_{\text{sys}}$ is defined in Eq.~\eqref{eq:mass-matrix}:

\begin{align}\label{eq:mass-matrix}
    M_{\text{sys}}=
    \begin{bmatrix}
         m_{\text{T}} & 0 & M_{13} & 0\\
         0 & m_{\text{T}} & -M_{26} & 0\\
         M_{13} & -M_{26} & M_{55} & 0\\
         0 & 0 & 0 & I_{\text{r}x}
    \end{bmatrix}
\end{align}

where $m_{\text{T}}$ is the total turbine mass and $I_{\text{r}x}$ is rotor inertia. Additional model parameters are defined as:

\begin{align}
    \begin{array}{lcl}\label{eq:mass-components}
        m_{\text{T}}=m_{\text{p}}+m_{\text{t}}+m_{\text{nc}}+m_{\text{r}} \\
        M_{13}=D_{\text{r}}(m_{\text{r}}+m_{\text{nc}})+m_{\text{t}} D_{\text{t}} \\
        M_{26}=d_{\text{nc}} m_{\text{nc}}-d_{\text{r}} m_{\text{r}} \\
        M_{55}=I_{\text{T}_y}+m_{\text{r}}(D_{\text{r}}^2+d_{\text{r}}^2)+m_{\text{nc}}(D_{\text{r}}^2+d_{\text{nc}}^2)+D_{\text{t}}^2m_{\text{t}} \\
        I_{\text{T}_y}=I_{\text{p}y}+I_{\text{t}y}+I_{\text{nc}y}+I_{\text{r}y},
    \end{array}
\end{align}

where $m_{\text{p}}$ is platform mass, $m_{\text{t}}$ is tower mass, $m_{\text{nc}}$ is nacelle mass, $m_{\text{r}}$ is rotor mass, $I_{\text{p}y}$ is platform inertia, $I_{\text{t}y}$ is tower inertia, $I_{\text{nc}y}$ is nacelle inertia, and $I_{\text{r}y}$ is rotor inertia. The dynamic equation given in Eq.~\eqref{eq:dynamics} is vectorized to efficiently calculate the time-domain the simulation. $\bar{A}$ and $\bar{C}_A$ in Eq.~\eqref{eq:dynamics} are defined as:

\begin{align}
     \bar{A} &=
    \begin{bmatrix}
         A_{11} & 0 & A_{13} & 0\\
         0 & A_{22} & 0 & 0\\
         A_{13} & 0   & A_{33} & 0\\
         0 & 0 & 0 & 0
    \end{bmatrix} \label{eq:hydro-added-mass-matrix}
    \\
    \bar{C}_A &=
    \begin{bmatrix}
        0 & 0 & A_{11}v_z & 0\\
        0 & 0 & -A_{11}v_x & 0\\
        -A_{11}v_z & A_{11}v_x & 0 & 0
    \end{bmatrix} \label{eq:hydro-coriolis-centripetal-matrix}
\end{align}

The parameters used in Eqs.~\eqref{eq:hydro-added-mass-matrix} and \eqref{eq:hydro-coriolis-centripetal-matrix} are defined as:

\begin{align}
    \begin{array}{lcl}\label{eq:hydro-added-mass-components}
        A_{11}=C_{\text{am}}\rho V_{\text{d}} \\ 
        A_{22}=C_{\text{am}}\left(\frac{1}{12}\rho\pi d_1^3\right) \\
        A_{13}=A_{11}(a_{\text{cv}}-a_{\text{pf}}) \\
        A_{33}=C_{\text{am}} I_{\text{add}}
    \end{array}
\end{align}
where $C_{\text{am}}$ is the hydrodynamic added mass coefficient, $\rho$ is water density, $V_{\text{d}}$ is submerged volume, $d_1$ is the platform base diameter, $a_{\text{cv}}$ is the distance between the platform base and the submerged volume center, $a_{\text{pf}}$ is distance between the platform Center of Gravity (COG) and the platform base, and $I_{\text{add}}$ is the inertia of the displaced volume.

Gravitational forces and moment are defined in Eq.~\eqref{eq:gravity-force-moment}, where $M^{\mathrm{m}}_{1\times 1}=g \sin{\theta_p}\left( D_r(m_{nc}+m_r)+D_tm_t \right)+g \cos{\theta_p}(d_{nc}m_{nc}-d_rm_r)$:

\begin{align}
    \begin{bmatrix}\label{eq:gravity-force-moment}
         F^{\mathrm{m}}_{2\times 1}\\
         M^{\mathrm{m}}_{1\times 1}\\
         0
    \end{bmatrix}_{4\times 1}&=
    \begin{bmatrix}
         m_Tg \sin{\theta_p}\\
         -m_Tg \cos{\theta_p}\\
         M^{\mathrm{m}}_{1\times 1}\\
         0
    \end{bmatrix}
\end{align}

and other external forces and moments are shown in Eq.~\eqref{eq:external-force}:

\begin{align}
    \begin{bmatrix}\label{eq:external-force}
         F_{2 \times 1}\\
         M_{1\times 1}\\
         \tau_{\mathrm{a}}-\tau_{\mathrm{g}}
    \end{bmatrix}_{4\times 1}&=
    \begin{bmatrix}
         F_{2\times1}^{\mathrm{hs}} + F_{2\times1}^{\mathrm{a}} + F_{2\times1}^{\mathrm{moor}} + F_{2\times1}^{\mathrm{hd}}\\
         M_{1\times1}^{\mathrm{hs}} + M_{1\times1}^{\mathrm{a}} + M_{1\times1}^{\mathrm{moor}} + M_{1\times1}^{\mathrm{hd}}\\
         \tau_{\mathrm{a}}-\tau_{\mathrm{g}}
    \end{bmatrix}
\end{align}
where $\theta_p$ is platform pitch, $F^{\mathrm{hs}}$ is the hydrostatic force, $F^{\mathrm{a}}$ is the aerodynamic force, $F^{\mathrm{moor}}$ is the mooring force, $F^{\mathrm{hd}}$ is the hydrodynamic force, $M^{\mathrm{hs}}$ is hydrostatic moment, $M^{\mathrm{a}}$ is the aerodynamic moment, $M^{\mathrm{moor}}$ is the mooring moment, and $M^{\mathrm{hd}}$ is the hydrodynamic moment.  These forces are calculated based on \citet{al2017dynamics} and are not detailed here for brevity.

\begin{figure}[ht!]
    \centering
    \includegraphics[width=1.8in]{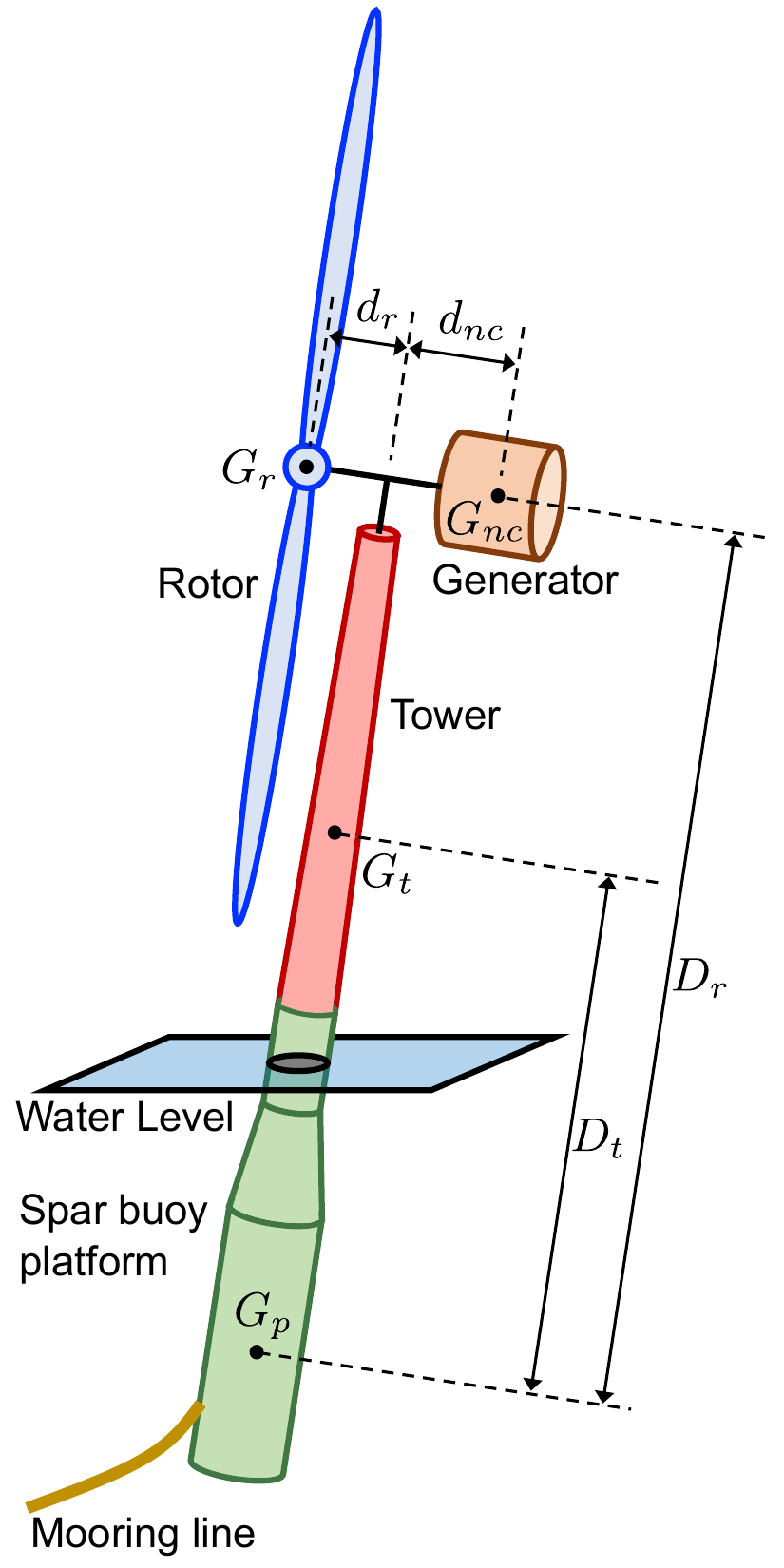}
    \caption{Schematic of the complete spar buoy-based FOWT system, including spar platform, mooring system, tower, generator, and rotor.}
    \label{fig:complete-system}
\end{figure}

\begin{figure}[ht!]
    \centering
    \subcaptionbox{}{\includegraphics[width=3cm]{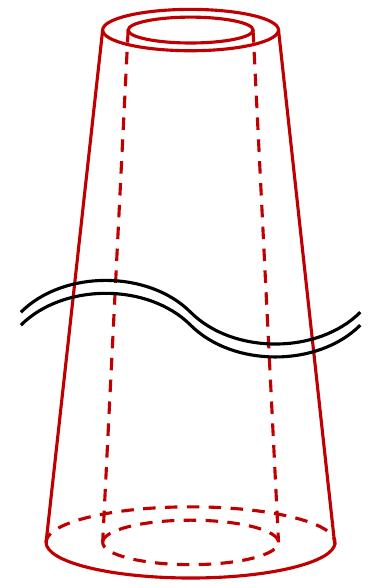}}
    \subcaptionbox{}{\includegraphics[width=4cm,trim={-0.4cm 0 0.4cm 0}]{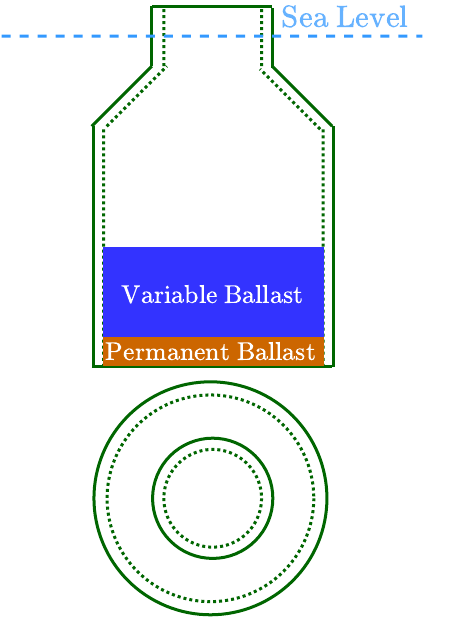}}
    \subcaptionbox{}{\includegraphics[width=3.6cm,trim={0 0 -1.5cm 0}]{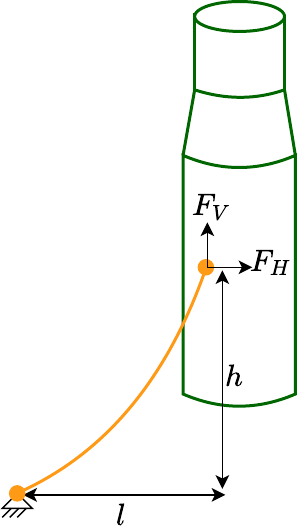}}
    \subcaptionbox{}{\includegraphics[scale=0.5]{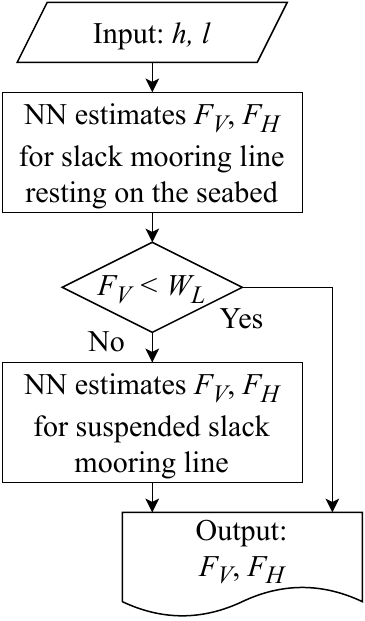}}
    \caption{Schematics of system components, including: (a) tower structure, (b) spar buoy platform, and (c) mooring line and a depiction its computational model (d) that uses a neural network (NN) to model the horizontal and vertical forces exerted by the mooring line.}
    \label{fig:system-components}
\end{figure}

\begin{figure}[ht!]
    \centering
    \subcaptionbox{}{\includegraphics[width=3.2cm,trim={0 0 -1cm 0}]{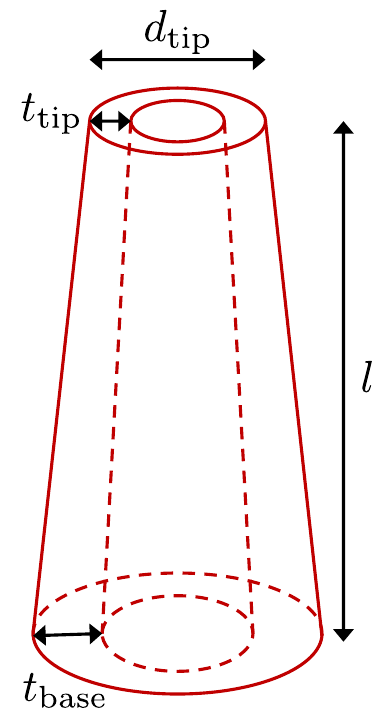}}
    \subcaptionbox{}{\includegraphics[width=8cm]{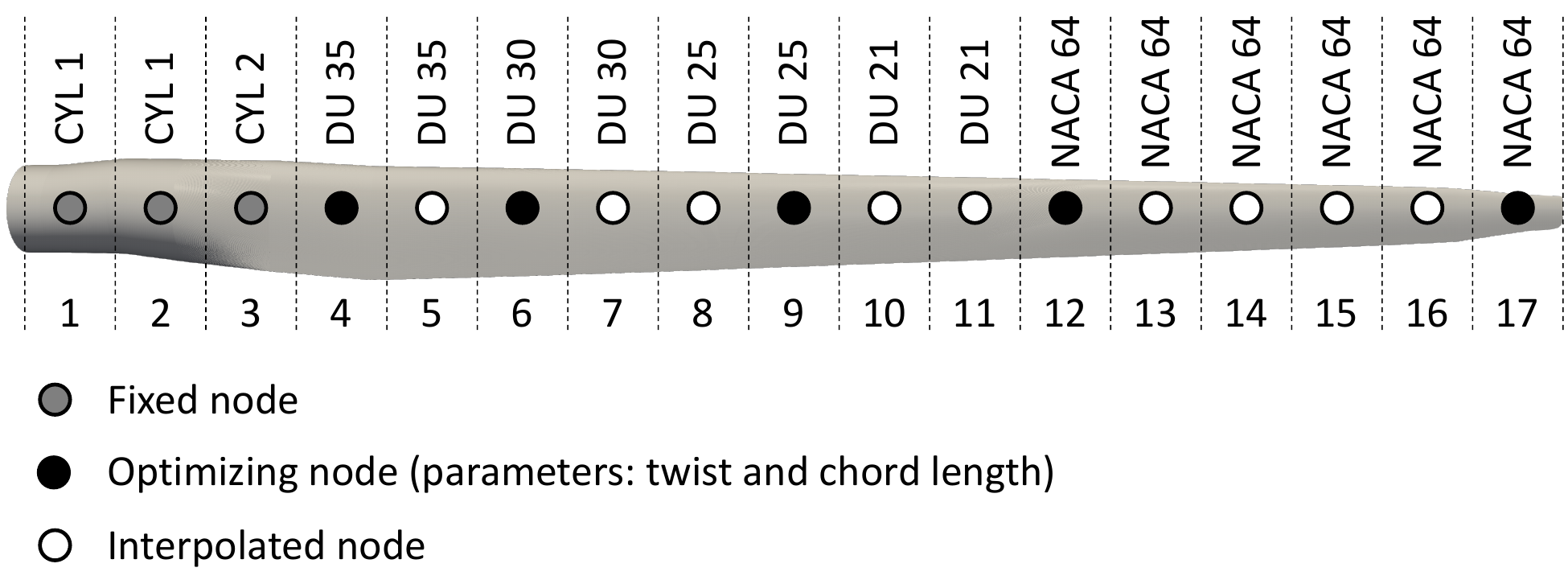}}
    \caption{Schematics illustrating plant design variables: (a) tower design variables; (b) blade design variables. Distributed blade design is parameterized using twist ($\phi$) and chord length ($\zeta$) values at each indicated `optimizing node'.}
    \label{fig:plant-parameters}
\end{figure}

\begin{figure*}[ht!]
    \centering
    \includegraphics[width=\textwidth]{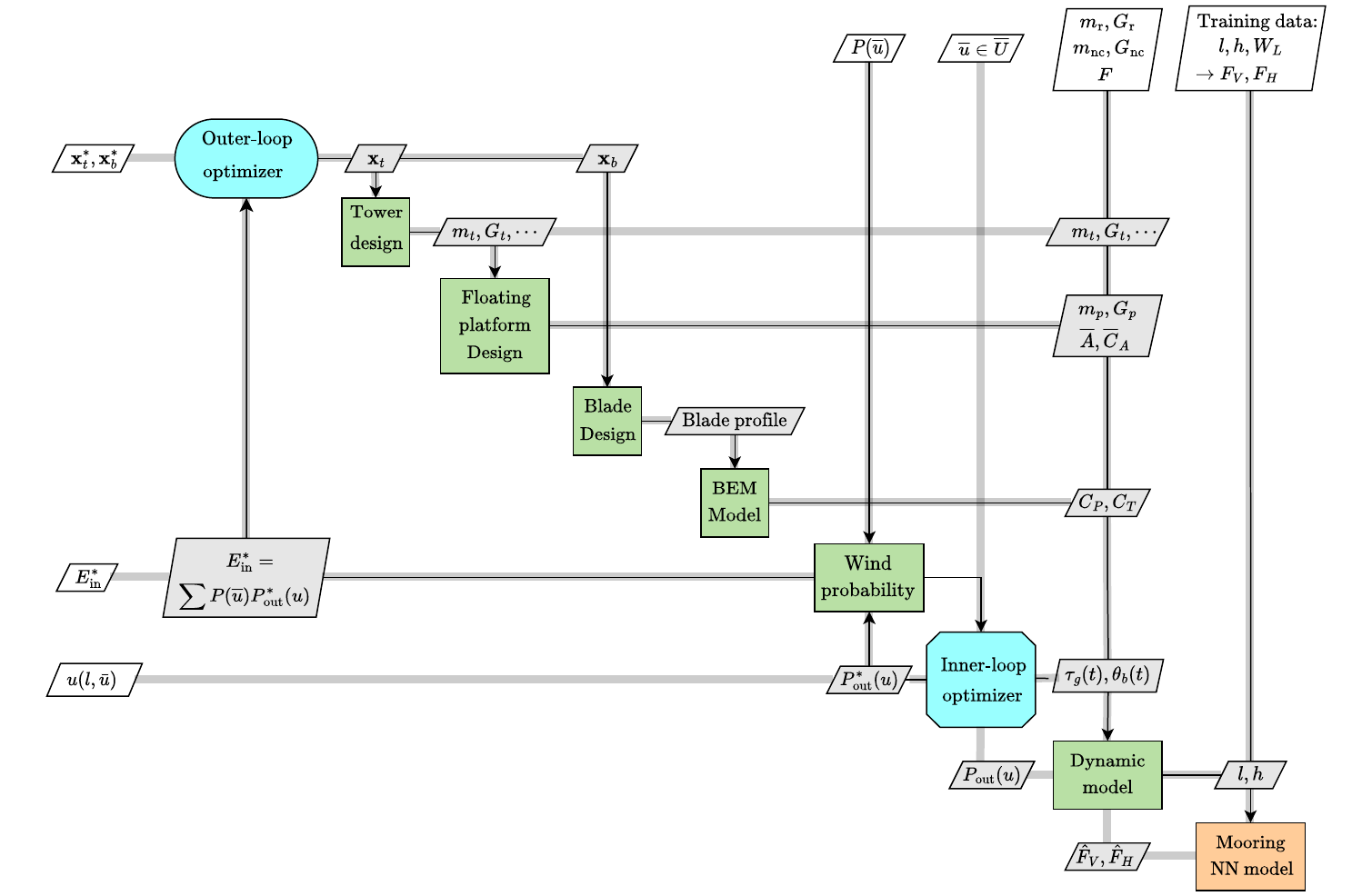}
    \caption{XDSM of the nested CCD problem depicting the solution process for the overall FOWT CCD problem. The inner-loop problem solves for the state and control trajectories, while the outer-loop problem solves for the plant design variables.}
    \label{fig:nested-ccd-flowchart}
\end{figure*}

Figure~\ref{fig:complete-system} shows a schematic of the complete spar buoy FOWT system, including all the system components considered in this study. In the figure, $d_{\text{r}}$ is the distance between the rotor COG and tower center line, $d_{\text{nc}}$ is the distance between the nacelle COG and the tower center line, $D_{\text{t}}$ is the distance between the tower and platform COGs, and $D_{\text{r}}$ is the distance between nacelle and platform COGs.

The complete system model presented in Fig.~\ref{fig:complete-system} comprised of the tower, spar buoy floating platform, slack mooring lines, rotor, and generator models. Figure~\ref{fig:system-components} illustrates the internal structure of the FOWT components considered in this study. Figure~\ref{fig:system-components}(a) depicts how the FOWT tower geometry is modeled as a tapered hollow cylinder (aspect ratio is exaggerated for visualization), and Fig.~\ref{fig:system-components}(b) shows the internal fixed and variable (water) ballast structures of the spar buoy platform. Figures \ref{fig:system-components}(c) and (d) depict the configuration and computing algorithm for the mooring system. In this study, we implemented and trained a multilayer perceptron neural network (NN) using \texttt{newff} and \texttt{train} MATLAB commands to calculate mooring line dynamics, instead of directly solving costly nonlinear equations. In Fig.~\ref{fig:system-components}(c), $l$, and $h$ are horizontal and vertical distances between the anchor point and the fairlead, $F_{\text{moor},H}$ is the horizontal mooring force, $F_{\text{moor},V}$ is the vertical mooring force, and $W_{\text{moor},L}$ is the line weight. It is required to use a different set of nonlinear equations to obtain the mooring forces depending on whether the mooring line is suspended or resting on the seabed \citep{al2017dynamics}. When vertical force is greater than mooring line weight, the mooring line is suspended; otherwise, it rests on the seabed. The neural network is trained for each distinct case. Inputs of the neural network include the relative position between the fairlead and the anchor point, and the outputs are the vertical and the horizontal mooring forces.

Figure~\ref{fig:plant-parameters} presents schematics that help define plant design variables. In this study the tower and blade designs are the extent of physical system design decisions. Figures \ref{fig:plant-parameters}(a) and (b) illustrate the tower and blade design variables, respectively. We defined four plant design variables for the tower: tower thickness at the base $t_{\mathrm{base}}$, tower thickness at the tip $t_{\mathrm{tip}}$, tower outer diameter at the tip $d_{\mathrm{tip}}$, and tower length $l$. We also defined ten blade design variables. The blade is divided into 17 nodes in the blade length-wise direction. Airfoil shape is predetermined for each blade node, and is not subject to design in this study. Distributed blade geometry may be adjusted by changing the values of blade twist ($\phi$) and chord length ($\zeta$) for select control nodes. These select nodes are depicted as `optimizing nodes' in  Fig.~\ref{fig:plant-parameters}(b)), shown as black circles. The first three nodes (gray nodes in Fig.~\ref{fig:plant-parameters}(b)) are assigned fixed values for twist and chord length. Twist and chord length values for interpolated nodes (white nodes in Fig.~\ref{fig:plant-parameters}(b)) are calculated by interpolating functions based on a B{\'e}zier curve obtained by fixed and optimized node values. Coordinates of the referenced airfoil shape profiles (CYL, DTU, and NACA) in Fig.~\ref{fig:plant-parameters}(b) can be found in the definition of NREL 5 MW floating offshore wind turbine \citep{Jonkman2009NREL5MW}.

Figure~\ref{fig:nested-ccd-flowchart} shows an extended design structure matrix (XDSM) of the nested CCD optimization process. The objective function of the overall problem is maximizing energy production, given as $E_{\text{in}}^{*}$. The tower and the blade design variables are represented as $\mathbf{x}_{\text{t}} = [ t_{\text{tip}},\; d_{\text{tip}},\; t_{\text{base}},\; l ]^{\textsf{T}}$ and $\mathbf{x}_{\text{b}} = [ \phi_{5 \times 1}^{\textsf{T}},\; \zeta_{5 \times 1}^{\textsf{T}} ]^{\textsf{T}}$, respectively. The superscripts $\dagger$ and $*$ represent the updated and optimal values, respectively. In the outer loop, the CMA-ES algorithm is used as an optimizer to iteratively create and update the plant design variables. This algorithm was found to handle the noisy response of the inner-loop problem efficiently and effectively. Power and thrust coefficients ($C_P$ and $C_T$) were calculated using blade element momentum (BEM) theory \citep{Hansen2013BEM}. For each plant design the spar buoy platform ballast mass required to achieve equilibrium in the heave direction is calculated. 

The inner-loop optimization problem is solved using the composition of these plant-dependent parameters ($C_P$ and $C_T$),  the dynamic model defined above, and the mooring force NN model. For each candidate plant design specified by the outer-loop plant design problem, the inner-loop problem is solved, and the resulting optimal control trajectories and the inner-loop objective function values for varied average wind speed points are sent back to the outer-loop optimizer.

\subsection{Objective and Constraint Functions}
\label{sec-sub:obj-and-constr}

The objective function used here is to maximize the AEP. To achieve this goal at the system-level, both the plant and control design variables need to be concurrently optimized. In this study, the plant design variables influence tower and blade geometry, and the control design variables are the time-dependent generator torque and blade pitch trajectories. The goal is to maximize the energy generated over a one-year long time horizon in the inner-loop optimization problem. The annual power generation is calculated based on a weighted sum of the sample simulations using a probabilistic wind distribution. Each sample simulation consists of 100 seconds of a predefined wind speed profile at varied average wind speed values following the probabilistic distribution by setting the initial time, $t_i$, as 0 and the final time, $t_f$, as 100 seconds in the inner-loop optimal control problems. Using this weighted sum method, the optimal design (in terms of both plant and control designs) converges to one that extracts the maximum average power for wind speeds with higher probability of occurrence. In contrast, the goal of maximum power can be relaxed for wind speeds with lower probability of occurrence.

The AEP can be obtained using Eq.~\eqref{eq:aep}:

\begin{align}
\begin{split}
\label{eq:aep}
\text{AEP} = 8,760 \int_{0}^{\infty} P_{\text{out}}(U) f(\bar{U}) \, \mathrm{d}\bar{U} \\
\approx 8,760 \sum_{j=1}^{N_{\text{B}}} P_{\text{out}}(u_j) P(\bar{u}_j)
\end{split}
\end{align}

where 8,760 is the total number of hours in a year, $\bar{U}$ is the nominal free wind velocity perpendicular to the rotor swept area, $U$ is the free wind speed at the hub height, $P_{\mathrm{out}}$ is the time-averaged wind turbine output power, and $f(U)$ is the Weibull probability density function. The free wind speed $U$ is also a function of tower length since the height of the tower determines the hub height. The nominal free wind speed, $\bar{U}$, is based on the free stream velocity data at the hub height (90 meters) for the NREL 5 MW reference turbine. The integral that calculates the AEP can also be transcribed to a summation over $N_{\text{B}}$ wind speed bins. $\bar{u}_j$ and $u_j$ are $j$-th discretized quantities of $\bar{U}$ and $U$ among $N_{\text{B}}$ wind speed bins, respectively.
\begin{figure}[ht!]
    \centering
    \includegraphics[width=6.8cm]{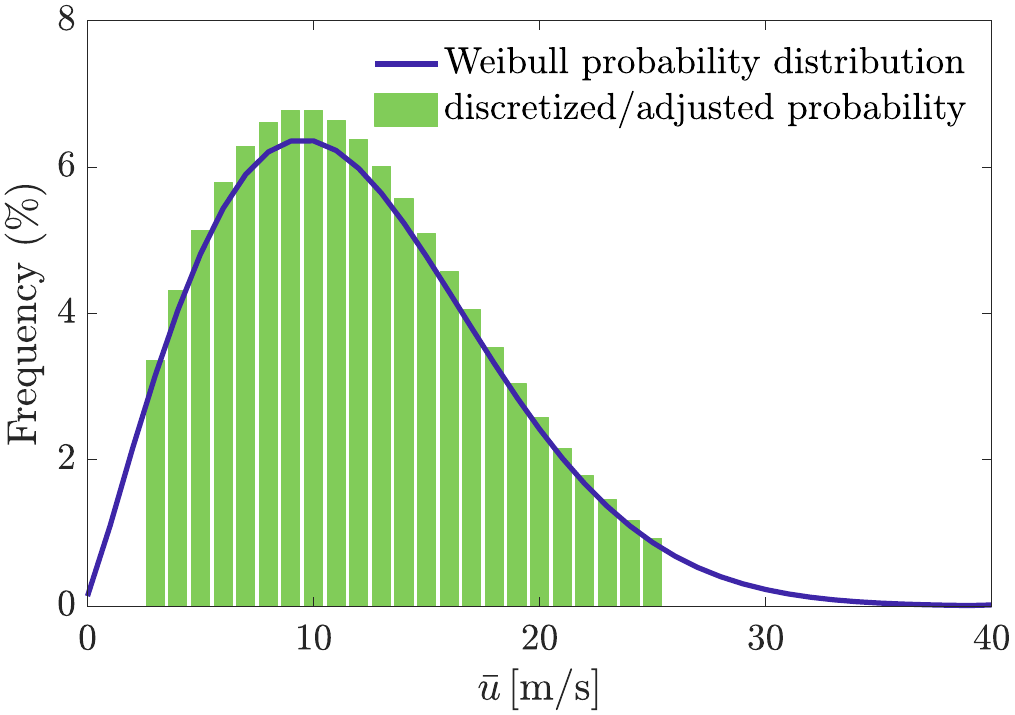}
    \caption{Free wind velocity values and the corresponding frequency of occurrence based on the Weibull probability distribution.}
    \label{fig:weibull-dist}
\end{figure}

\begin{figure}[ht!]
    \centering
    \subcaptionbox{}{\includegraphics[width=7.1cm]{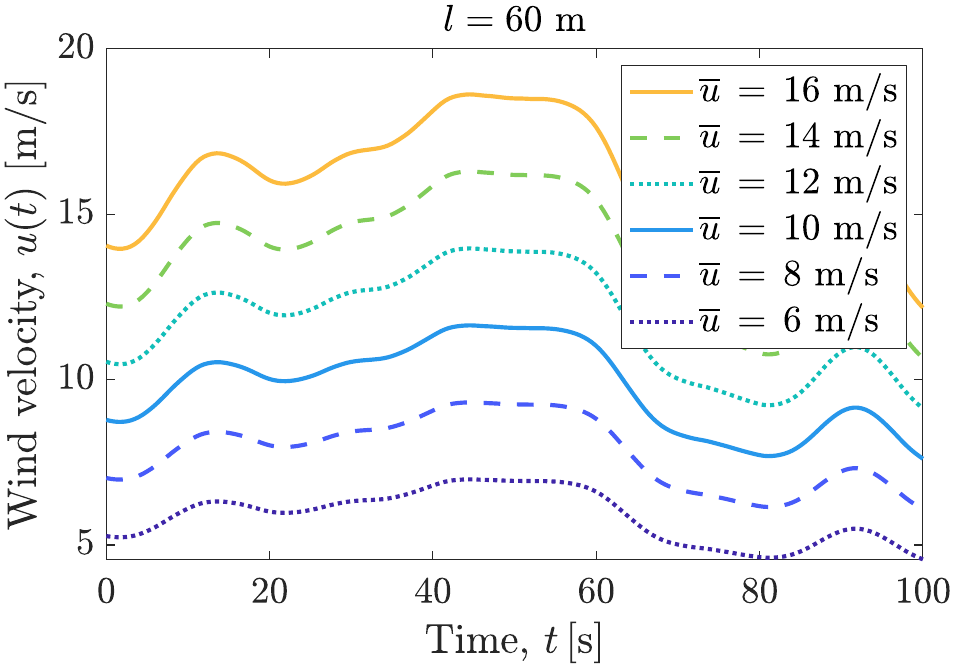}} \;\;\;\;
    \subcaptionbox{}{\includegraphics[width=7.1cm]{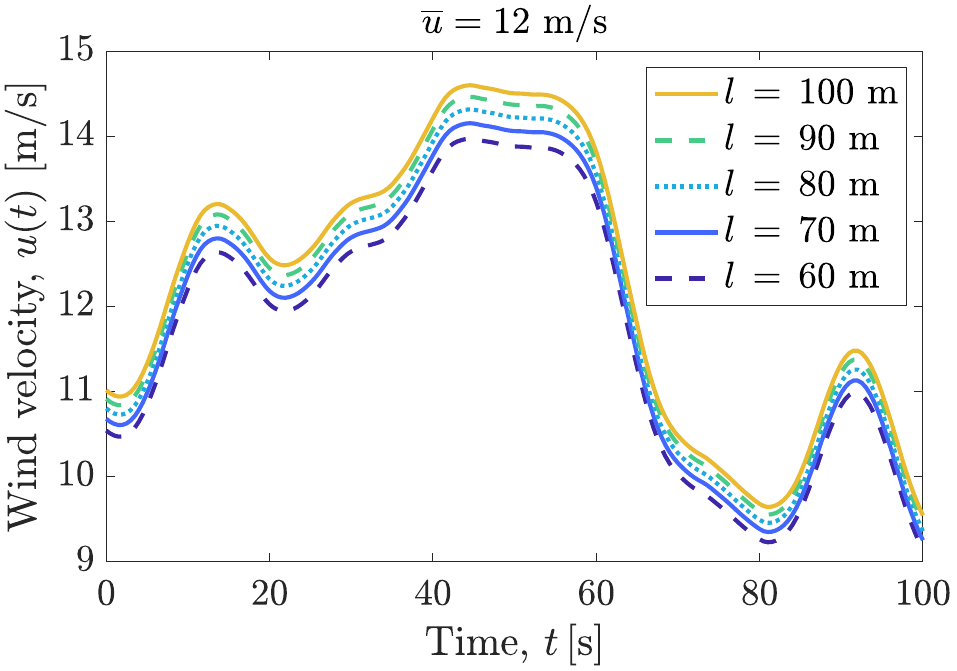}}
    \caption{Wind profiles in 100 seconds of time horizon for varied mean wind velocity ($\bar{u}$) and varied tower height ($l$). (a) Wind profiles for mean wind velocity values ranging from 6 to 16 m/s; (b) Wind profile for mean wind velocity of 12 m/s and tower height values ranging from 60 to 100 m.}
    \label{fig:wind-function}
\end{figure}

Here we use Weibull probability distribution as a wind speed probabilistic occurrence model over one year. The probability density function of the Weibull distribution is shown in Eq.~\eqref{eq:weibull-dist}:

\begin{align}\label{eq:weibull-dist}
    & f\left(\bar{U}\right)=\frac{k}{c}\left(\frac{\bar{U}}{c}\right)^{k-1}\exp\left(-\left(\frac{\bar{U}}{c}\right)^k\right)
\end{align}
where $k > 0$ is the shape parameter (here, $k = 2$), and $c > 0$ is the scale parameter of the distribution (here $c = 13.44$). This wind speed probabilistic occurrence model is discretized with an interval of 1 m/s over a 3 to 25 m/s range and adjusted to have a sum of 100\% for its discretized range, resulting in $N_B=23$ wind speed bins . The probability density function is illustrated in Fig.~\ref{fig:weibull-dist}. The wind load applied at the center of the tower is also compensated by the power law to account for wind shear, shown in Eq.~\eqref{eq:wind-power-law}: 

\begin{align}\label{eq:wind-power-law}
    u = u\left( \bar{u},l \right)=\bar{u}\left(\frac{l+z_{\text{base}}}{l_{\text{baseline}}+z_{\text{base}}}\right)^{0.2}
\end{align}
This accounts for the tower height $l$ variation, where $l_{\text{baseline}}$ is the height of the NREL 5 MW baseline wind turbine tower (76 meters) and $z_{\text{base}}$ is the distance from sea mean water level to tower base, which is 12.4 meters. Figure~\ref{fig:wind-function}(a) shows the wind velocity profiles for select $\bar{u}$ values with the baseline tower length, and Fig.~\ref{fig:wind-function}(b) shows the wind velocity profiles as a function of tower length with a specific $\bar{u}$ value of 12 m/s. Increased tower length tends to exhibit increased AEP in general because the free wind velocity at the higher altitude is larger, as shown in Eq.~\eqref{eq:wind-power-law}. However, this trend only holds up to a specific limit in practice because taller towers are also subject to increased thrust force, which leads to an increased platform pitching motion. Increased platform pitch amplitudes will eventually reach the pitch limits enforced by path constraints in the inner loop optimal control problem. Excessive platform pitch can be curtailed in this CCD optimization problem through less aggressive generator torque or blade pitch control. This limits AEP gains with tower height. In other words, at some value for tower height, platform pitch limits AEP gains. Broader design changes, such as those to the platform or mooring subsystems, could help increase the tower height at which platform pitch begins to constrain AEP.

Optimization problems are often posed as minimization problems for compatibility with established optimization solvers and by convention. The outer loop objective function is defined as:

\begin{align}
    &J_{\text{out}} = -8,760 \sum_{j=1}^{N_B} P\left(\bar{u}\right) P_{\text{out}} \left(u\right) = -8,760\cdot E_{\text{in}}\label{eq:J-out}\\
    &\text{where: } E_{\text{in}} = \sum_{j=1}^{N_B} P\left(\bar{u}\right) P_{\text{out}}\left(u\right)\label{eq:E-in}
\end{align}
$E_{\text{in}}$ is obtained by solving the inner-loop problem across 23 wind speeds with their corresponding probabilities, as shown in Fig.~\ref{fig:weibull-dist}. The objective of each inner-loop optimization problem is given in Eq.~\eqref{eq:J-in}:
\begin{equation}\label{eq:J-in}
    J_{\text{in}} = - (t_f - t_i) P_{\text{out}}\left(u\right)
\end{equation}
The time-averaged output power $P_{\text{out}}$ for each OLOC problem is computed by Eq.~\eqref{eq:P-out}:
\begin{equation} \label{eq:P-out}
    P_{\text{out}}\left(u\right) = \frac{1}{t_f - t_i} \int_{t_i}^{t_f} \left(P_a- 10^{-7}\dot{\tau}_g^2-10^7\dot{\theta}_b^2\right) dt
\end{equation}
where $P_{\text{a}}$ is aerodynamic power, $\dot{\tau}_{\text{g}}$ is generator torque rate, and $\dot{\theta}_{\text{b}}$ is blade pitch rate. The generator torque and blade pitch rates are control trajectories that are optimized in the inner-loop problem. The inner-loop objective function includes control cost terms (i.e., small penalties on the torque and blade pitch rates), to produce a non-singular optimal control problem and prevent bang-bang or singular arc control trajectories in the solution. Also, as discussed in \citet{gros2013economic}, the aerodynamic power is used as an integrand instead of the generator power to prevent turnpike, which is embodied in turbine problems as the tendency to maximize generator torque at the termination of a finite time horizon. Parameter bounds and path constraint functions for the inner-loop optimization problem are defined in Tables~\ref{tab:bound-constr} and \ref{tab:path-constr}. The results given in Section \ref{sec:results} indicate that some of these constraints are active in certain regions of operation. Once all the inner-loop problems (in this study, 23 inner-loop OLOC problems) are solved, $E_{\text{in}}$ is updated by using Eq.~\eqref{eq:E-in} for computing the outer-loop objective function, given in Eq.~\eqref{eq:J-out}.

\begin{table}[ht!]
\centering
\caption{Parameter bounds for the inner-loop OLOC problem.}
\label{tab:bound-constr}
\resizebox{0.49\textwidth}{!}{ 
\begin{tabular}{llrrr}
\toprule
                 &                           &            & Lower    & Upper  \\
Name             & Explanation               & Unit       & bound    & bound  \\
\midrule
$\omega_r$       & rotor rotational velocity & rad/s      & $0.00$   & 1.51   \\
$\theta_p$       & platform pitch            & deg        & $-6.3$   & 6.3    \\
$\theta_b$       & blade pitch               & rad        & $0.0$    & 40.0   \\
$\tau_g$         & generator torque          & $10^6$ N-m & $0.00$   & 4.18   \\
$\dot{\theta}_b$ & blade pitch rate          & rad/s      & $-0.57$  & 0.57   \\
$\dot{\tau}_g$   & generator torque rate     & MW         & $-0.1$   & 0.1    \\
$\phi$           & blade twist               & deg        & $-0.001$ & 15.970 \\
$\zeta$          & blade chord length        & m          & $0.014$  & 5.580  \\
\bottomrule
\end{tabular}
}
\end{table}

\begin{table}[ht!]
\centering
\caption{Path constraints for the inner-loop OLOC problem.}
\label{tab:path-constr}
\resizebox{0.49\textwidth}{!}{ 
\begin{tabular}{llrr}
\toprule
Name      & Explanation                 & Unit  & Constraint \\
\midrule
$\sigma$  & Tower maximum static stress & MPa   & $\sigma \le 45.0$ or $90.0$  \\ 
$P_u$     & Generator power             & MW    & $P_u \le 5.0$      \\
\bottomrule
\end{tabular}
}
\end{table}

In the nested CCD strategy introduced previously in Fig.~\ref{fig:nested-ccd-flowchart}, the outer loop optimizer updates the plant design. In this study, we use CMA-ES to generate a set of design points (population) to evaluate, and the covariant matrix adaptation procedure finds the direction toward the optimal solution. For the design points generated by the CMA-ES, the inner-loop OLOC problem is solved. The Weibull probabilistic distribution is utilized to divide wind speeds into 23 bins with center $\bar{u} \in \mathcal{U}$, where $\bar{u}$ represents the average wind velocity. As defined in Eq.~\eqref{eq:wind-power-law}, the wind speed for defining wind load $u$ is a function of $\bar{u}$ and $l$, and this relationship depends on tower height. The outer-loop objective function is equivalent to the weighted sum of the $N_B$ inner-loop objective functions, which is required for stable convergence of the nested CCD problem. When the outer-loop optimization problem converges, the CMA-ES algorithm provides the final optimal solution and will be terminated.

\section{Results and Discussion}
\label{sec:results}
This section comprises several studies aimed at showcasing the results of the CCD approach and gaining insights from different scenarios. In Section \ref{sec:Control_Co-Desig_With_Tower_Only_and}, two cases are discussed: 1) CCD with four tower parameters as plant design variables, and
2) CCD with four tower parameters and ten blade parameters as plant design variables. This section presents the obtained optimal plant variables and objective values. It also includes an illustration of the CMAES population, and a sensitivity study is performed. Section \ref{sec:omparison of Increased Maximum Allowable Tower Stress} investigates the impact of changing the maximum allowable tower stress constraint on power generation and Annual Energy Production (AEP). It should be noted that here, stress resulting from fore-aft bending was only considered. Throughout this paper, when discussing stress upper bound, we specifically mean the upper bound of fore-aft bending stress. This constraint is used here just as a means to observe the coupling between plant and controller, and to see how the design changes as we modify the upper bound. In future work, additional constraints and higher fidelity models will be employed. Section \ref{Comparison Studies Showing Tower Mass Effect on AEP} explores the effect of increasing tower mass on the time series trajectories within the inner loop. Lastly, Section \ref{sec:Wave_and_Fatigue_Study} demonstrates the inner loop trajectories in the presence of waves and includes a fatigue study. These sections collectively offer a comprehensive analysis of CCD results and provide valuable insights into various aspects of the design and optimization process of floating offshore wind turbines.

\begin{table}[ht!]
\centering
\caption{Optimal plant design variable values and corresponding merit function values for optimization with varied targets. The tower-only (T) CCD problem solves for tower design variables in conjunction with control trajectories. The tower and blades (T\&B) CCD problem solves for the tower design parameters, blade design parameters, and control trajectories. Values with gray shading represent parameters are not optimized and held constant.}

\label{tab:optimized-parameters}
\resizebox{0.49\textwidth}{!}{
\begin{tabular}{cccrrr}
    \toprule
    & & & \multicolumn{3}{c}{Optimization target} \\
    \cmidrule(){4-6}
    Group & Variable & Unit 
    & \hspace*{-0.5em}\shortstack[r]{None\\(baseline)}
    & \shortstack[r]{\;\;Tower\\only}
    & \shortstack[r]{Tower \&\\blades} \\
    \midrule
    \multirow{4}{*}{\shortstack[c]{Tower\\parameters}}
    & $t_{\text{base}}$    & m
    & \cellcolor[HTML]{cfcfcf} 0.027 & 0.042 & 0.042 \\
    & $t_{\text{tip}}$     & m
    & \cellcolor[HTML]{cfcfcf} 0.019 & 0.012 & 0.012 \\
    & $d_{\text{tip}}$     & m
    & \cellcolor[HTML]{cfcfcf} 3.870 & 4.95 & 5.00\\ 
    & $l$                  & m
    & \cellcolor[HTML]{cfcfcf} 77.600 & 83.04 & 81.38 \\
    \cmidrule(r){1-3}\cmidrule(){4-6}
    \multirow{10}{*}{\shortstack[c]{Blade\\parameters}}
    & $\phi_{4}$             & deg
    & \cellcolor[HTML]{cfcfcf} 13.31 & \cellcolor[HTML]{cfcfcf} 13.31 &  12.09 \\
    & $\phi_{6}$             & deg
    & \cellcolor[HTML]{cfcfcf} 11.48 & \cellcolor[HTML]{cfcfcf} 11.48 &  9.23 \\
    & $\phi_{9}$             & deg
    & \cellcolor[HTML]{cfcfcf} 6.54 & \cellcolor[HTML]{cfcfcf} 6.54 &  3.96 \\
    & $\phi_{12}$            & deg
    & \cellcolor[HTML]{cfcfcf} 1.53 & \cellcolor[HTML]{cfcfcf} 1.53 &  1.52 \\
    & $\phi_{17}$            & deg
    & \cellcolor[HTML]{cfcfcf} 0.11 & \cellcolor[HTML]{cfcfcf} 0.11 & 0.06 \\
    & $\zeta_{4}$             & m
    & \cellcolor[HTML]{cfcfcf} 4.557 & \cellcolor[HTML]{cfcfcf} 4.557 &  5.580 \\
    & $\zeta_{6}$             & m
    & \cellcolor[HTML]{cfcfcf} 4.007 & \cellcolor[HTML]{cfcfcf} 4.007 &  5.056 \\
    & $\zeta_{9}$             & m
    & \cellcolor[HTML]{cfcfcf} 3.502 & \cellcolor[HTML]{cfcfcf} 3.502 &  3.136 \\
    & $\zeta_{12}$            & m
    & \cellcolor[HTML]{cfcfcf} 2.764 & \cellcolor[HTML]{cfcfcf} 2.764 &  2.458 \\
    & $\zeta_{17}$            & m
    & \cellcolor[HTML]{cfcfcf} 1.419 & \cellcolor[HTML]{cfcfcf} 1.419 &  1.458 \\
    \cmidrule(r){1-3}\cmidrule(){4-6}
    \multirow{2}{*}{\shortstack[c]{Merit\\functions}}
    & $-J^{*}_{\text{out}}$ & GWh
    & 26.24 & 29.22& 29.34 \\
    & $m_{\text{tower}}$   & tonne
    & 249.6 & 344.4 & 342.0 \\
    \bottomrule
\end{tabular}}
\end{table} 

\begin{figure}[ht!]
    \centering
    \subcaptionbox{}{\includegraphics[width=4.2cm]{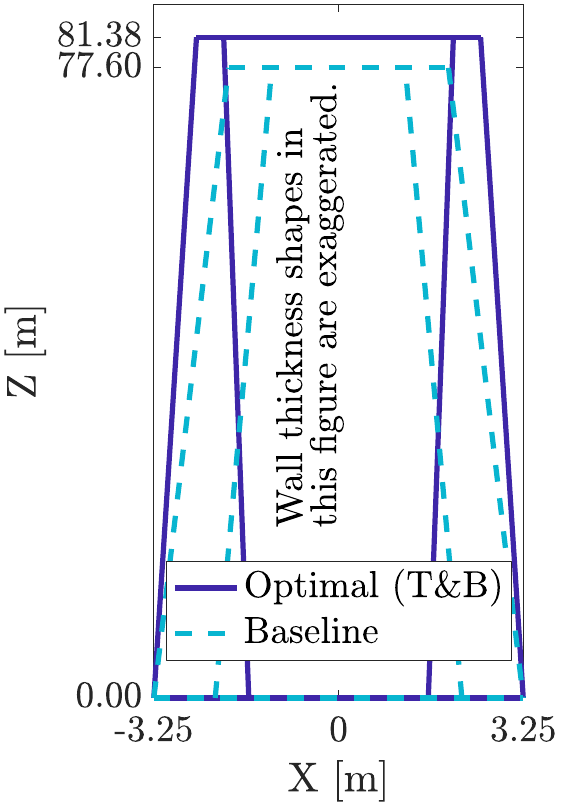}} \;\;\;\;
    \subcaptionbox{}{\includegraphics[width=7.3cm]{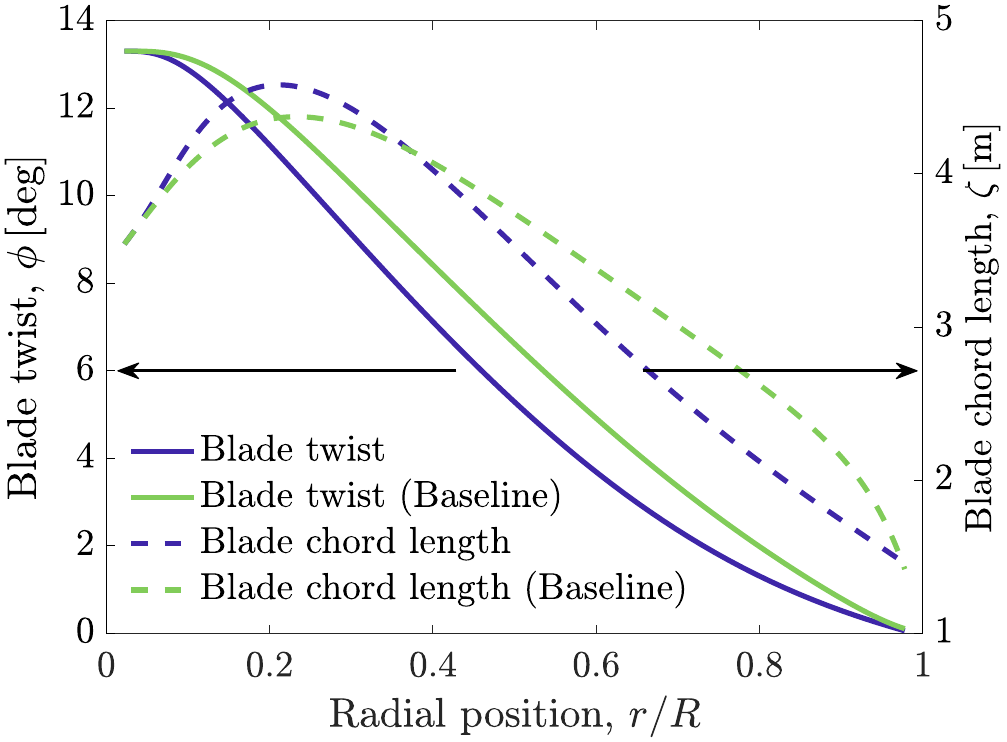}}
    \caption{Optimized tower and blade designs for the CCD problem with both tower and blade (T\&B) variables optimized. (a) Optimized tower shape compared to the baseline design; (b) Optimized blade twist ($\phi$) and chord length ($\zeta$) over normalized blade radial position ($r/R$). The tower wall thickness shapes are exaggerated for visual comparison. Readers are referred to `Tower \& blades' column in Table~\ref{tab:optimized-parameters} for accurate wall thickness values.}
    \label{fig:optimized-tower-shape}
\end{figure}

\begin{figure}[ht!]
    \centering
    \subcaptionbox{}{\includegraphics[width=7.6cm]{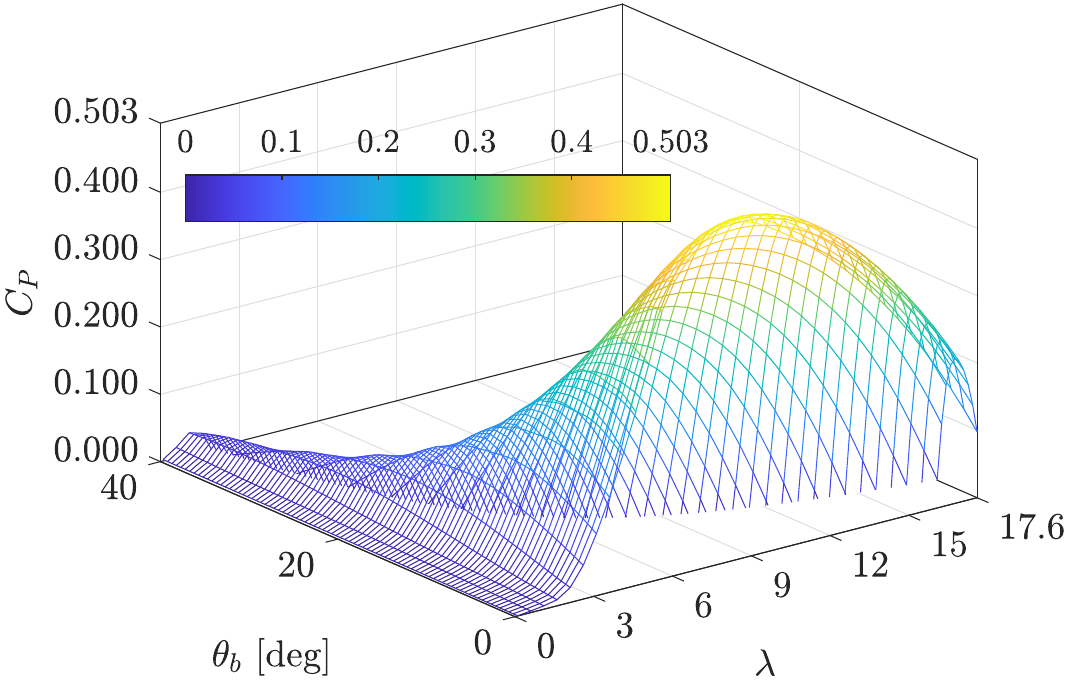}}
    \subcaptionbox{}{\includegraphics[width=7.6cm]{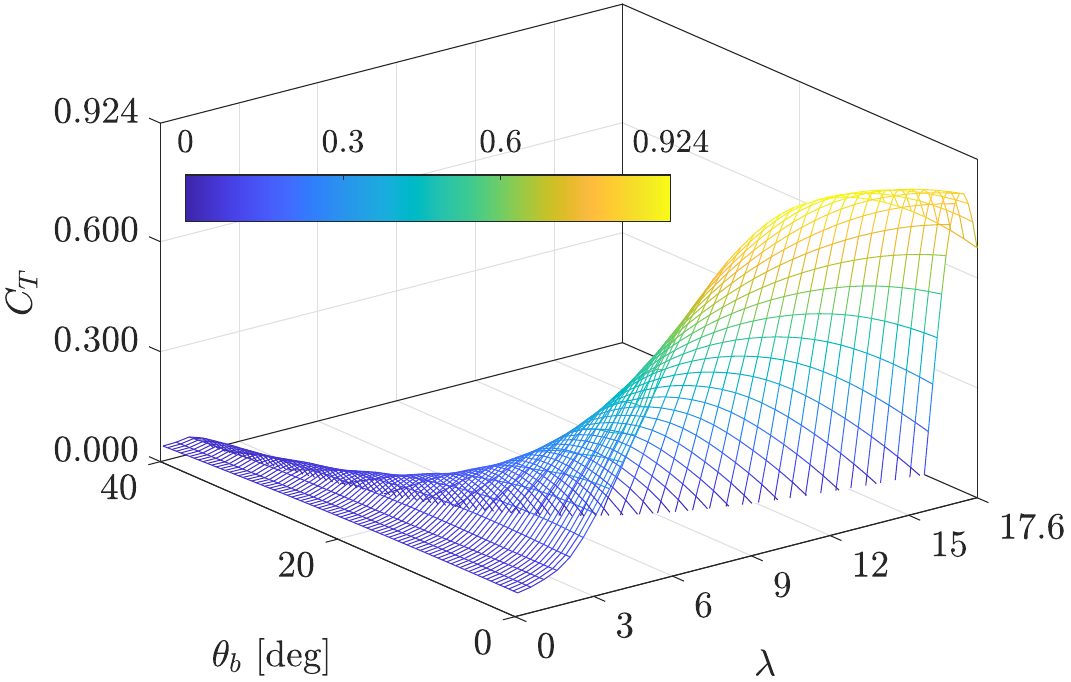}}
    \caption{Optimized blade power and thrust coefficients as functions of tip speed ratio ($\lambda$) and blade pitch angle ($\theta_{b}$). (a) Power coefficient surface; (b) thrust coefficient surface.}
    \label{fig:optimized-blade-coeffs}
\end{figure}

\subsection{Control Co-Design With Tower Only and Tower and Blade Plant Design Variables}
\label{sec:Control_Co-Desig_With_Tower_Only_and}

This section presents the results of two CCD studies and compares them with the optimal control result of the baseline NREL 5 MW wind turbine with the same model  presented in Section \ref{sec:prob-def}. Unless mentioned otherwise, the maximum tower stress constraint value used in this study is 45 MPa. The summary of the problems we compare in this study are listed in Table~\ref{tab:optimized-parameters}, where $\phi$ is blade twist, $\zeta$ is blade chord length, and indices represent  blade node values. Readers are also referred to Fig.~\ref{fig:plant-parameters} for the nodal information where design variables are located. Gray-shaded cells in the table show which design variables are held fixed and are not optimized during the CCD procedure. The `None (baseline)' column represents the NREL 5 MW baseline result. All plant design variables are shaded (fixed parameters), and the inner-loop OLOC problem is the only design problem that is solved in this case to generate the optimal AEP, $J_{\text{out}}^{\ast}$. The next `Tower only' column presents the results from the first CCD problem, which optimizes the tower design variables and the control trajectories. The last `Tower \& blades' column corresponds to the second CCD problem, which optimizes both the tower and blade design variables, as well as the control trajectories.

As we see from the overall AEP (objective function) values, $-J_{\text{out}}^{*}$, from Table~\ref{tab:optimized-parameters}, increased design flexibility with more plant design degrees of freedom help improve the objective function value significantly. Compared to the baseline design, optimizing the tower and blade design variables and the control trajectories together increases AEP by 7.17\%. Figure~\ref{fig:optimized-tower-shape}(a) depicts the optimized tower design for the CCD case with both the tower and blade design variables optimized. The converged design solution indicates that the tower height has increased to a particular height (here, 83.12 m) that is larger than the baseline tower height (77.60 m) design. In addition, the tower thickness has also increased to satisfy tower stress constraints. The optimized blade is illustrated in Fig.~\ref{fig:optimized-tower-shape}(b), and the resulting power and thrust coefficients corresponding to this optimal blade design are shown in Fig.~\ref{fig:optimized-blade-coeffs}. The power and thrust coefficients are functions of blade pitch $\theta_{\text{b}}$ and tip speed ratio $\lambda$.

\begin{figure}[ht!]
    \centering
    \subcaptionbox{}{\includegraphics[width=7.2cm]{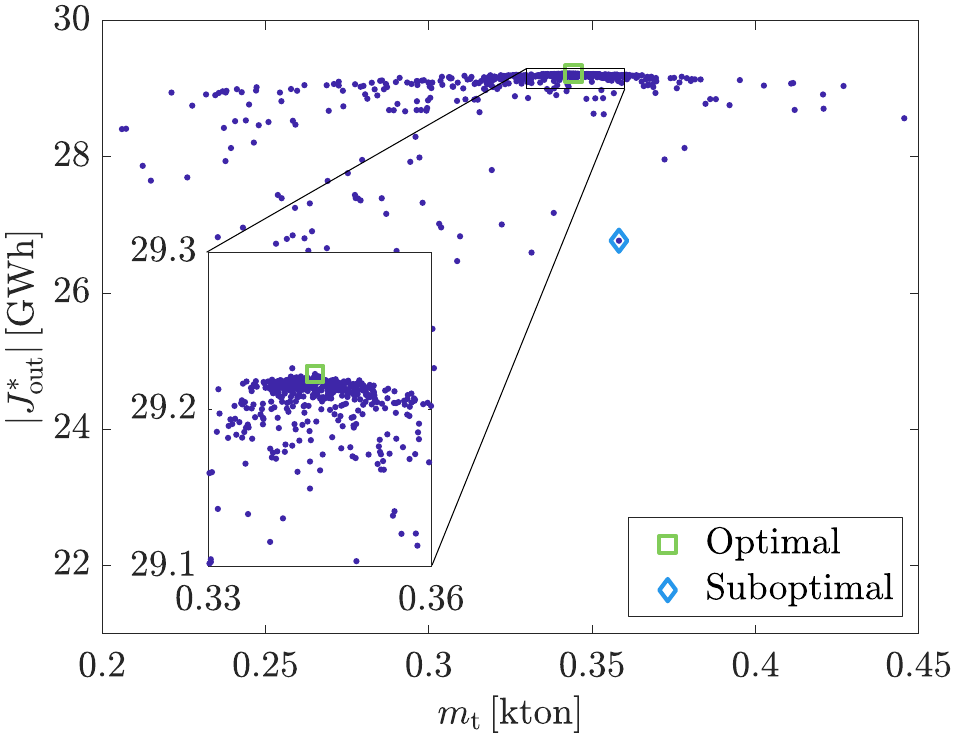}}
    \subcaptionbox{}{\includegraphics[width=7.2cm]{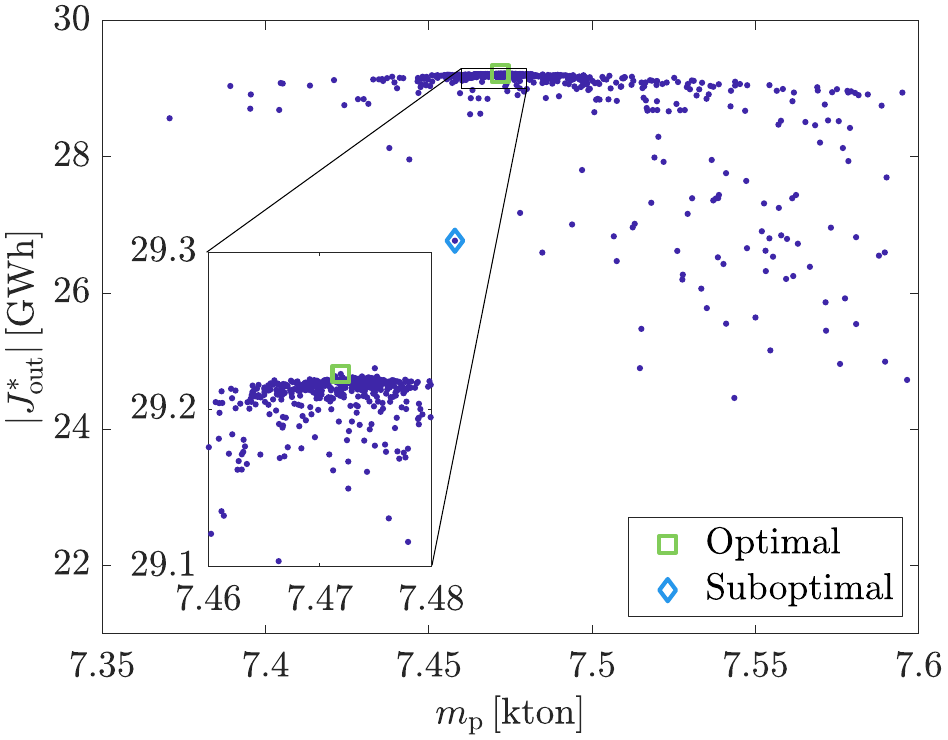}}
    \caption{Explored designs in the two-dimensional space of AEP and mass of FOWT components for the CCD problem with tower-only (T) plant optimization. Square markers represents the optimal plant and control design points. Diamond markers represent a selected suboptimal design point with a tower design that is heavier than the optimal design. Circular dots represent explored design points. Here, platform design is not optimized, but ballast mass is adjusted to keep the mean water level of the floating platform consistent. (a) Explored designs in AEP and tower mass space; (b) Explored designs in AEP and platform mass space.}
    \label{fig:aep-ptfmmass}
\end{figure}

\begin{figure}[ht!]
    \centering
    \includegraphics[width=4.2cm]{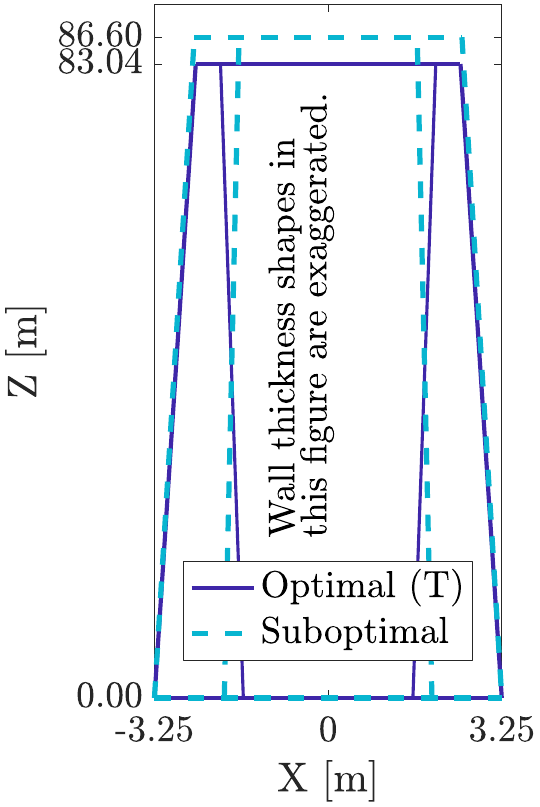}
    \caption{Resulting tower shape for the two design points highlighted in Fig.\ref{fig:aep-ptfmmass}. The selected suboptimal design point has a larger tower mass, but the AEP for the optimal design point (29.22~GWh) is about 9\% larger than the AEP for the suboptimal design point (26.77~GWh). }
    \label{fig:result-tower-shapes}
\end{figure}

Candidate designs that were explored during the CMA-ES optimization procedure are shown in Fig.~\ref{fig:aep-ptfmmass}, depicted in two distinct two-dimensional spaces. The corresponding tower shapes for the optimal design and a selected suboptimal design are shown in Fig.~\ref{fig:result-tower-shapes}. In Fig.~\ref{fig:aep-ptfmmass}, the optimal design point is indicated with a green square, and a sample suboptimal design point with a slightly larger tower mass is marked with a blue diamond. The indicated optimal design point exhibits the highest AEP value among all explored design points. The zoomed-in view of the plot shows that the optimal point has one of the largest objective function values among all of its neighbors. Across all of the designs explored in this study, the platform shape is fixed. As tower mass changes, the variable ballast (ballast water mass in practice) is adjusted to satisfy the force equilibrium requirement in the heave direction. Therefore, the platform capital cost for all these design points will be the same, but the tower cost varies with tower design changes.

Suppose we consider a multiobjective optimization problem for maximizing the AEP and minimizing the tower mass. This quantifies the best-possible tradeoffs between AEP and mass. In this case, all the design points located on the right side of the hypothetical vertical line crossing through the optimal design point are inferior to the optimal design since both performance indices (the AEP and the tower mass) are dominated by the optimal design. However, on the left side of the vertical line, finding good nondominated solutions in terms of superior tower mass value by sacrificing the AEP value is possible. Thus, we have identified a strategy for expanding this CCD problem into a multiobjective optimization study. Additional objective functions that consider additional performance and cost indices beyond AEP and mass may be used.

\begin{table}[ht!]
\centering
\caption{Sensitivity study results with $\pm 5\%$ perturbations of tower design variable values obtained from the tower-only CCD optimization result.}
\label{tab:sensitivity}
\resizebox{0.49\textwidth}{!}{
\begin{tabular}{cccccc}
\toprule
Tower design    & $\Delta_{\mathrm{rel}}$ & Value & $J_{\text{out}}$  & $\Delta J_{\mathrm{rel}}$ & $\Delta J_{\mathrm{rel}}/\Delta_{\mathrm{rel}}$ \\
parameter & [\%] & [m] & [GWh] & [\%] & [-] \\
\midrule
$t_{\text{base}}^{+}$   & $+5.0$  & 0.0445  & $-28.7839$  & $-1.49$  & $-0.298$  \\
$t_{\text{base}}^{-}$   & $-5.0$  & 0.0403  & $-29.2094$  & $-0.04$  & $-0.008$  \\
$t_{\text{tip}}^{+}$    & $+5.0$  & 0.0120  & $-29.2083$  & $-0.04$  & $-0.008$  \\
$t_{\text{tip}}^{-}$    & $-5.0$  & 0.0108  & $-26.9892$  & $-7.63$  & $-1.526$  \\
$d_{\text{tip}}^{+}$    & $+5.0$  & 5.353   & $-29.2076$  & $-0.04$  & $-0.008$  \\
$d_{\text{tip}}^{-}$    & $-5.0$  & 4.843   & $-29.2134$  & $-0.02$  & $-0.004$  \\
$l^{+}$                 & $+5.0$  & 86.751  & $-29.1929$  & $-0.09$  & $-0.018$  \\
$l^{-}$                 & $-5.0$  & 78.489  & $-29.2023$  & $-0.06$  & $-0.012$  \\
\bottomrule
\end{tabular}}
\end{table}

A sensitivity analysis has been performed with a $\pm 5\%$ perturbation of tower design variables based on the optimal design resulting from the tower-only CCD problem. The sensitivity results confirm the validity of our design solution by demonstrating performance reductions. Table~\ref{tab:sensitivity} presents a summary of the sensitivity analysis results. In this table, a relative perturbation $\Delta_{\text{rel}}$ for an arbitrary quantity $\chi$ is defined as $\Delta_{\text{rel}}=\left(100\%\right)(\chi-\chi^{\ast})/\chi^{\ast}$. Similarly, a relative change in AEP $\Delta J_{\text{rel}}$ is defined as $\Delta J_{\text{rel}}=\left(100\%\right)(\left|J_{\text{out}}\right|-\left|J^{\ast}_{\text{out}}\right|)/\left|J^{\ast}_{\text{out}}\right|$. As observed in the results, varying any tower design variable by 5\% in any direction from the optimal design decreases the AEP. In addition, the table indicates that the tower tip thickness and tower base thickness are the most sensitive parameters that affect the objective function value.

\subsection{Comparison of Increased Maximum Allowable Tower Stress}
\label{sec:omparison of Increased Maximum Allowable Tower Stress}

\begin{figure}[ht!]
    \centering
    \includegraphics[width=4.2cm]{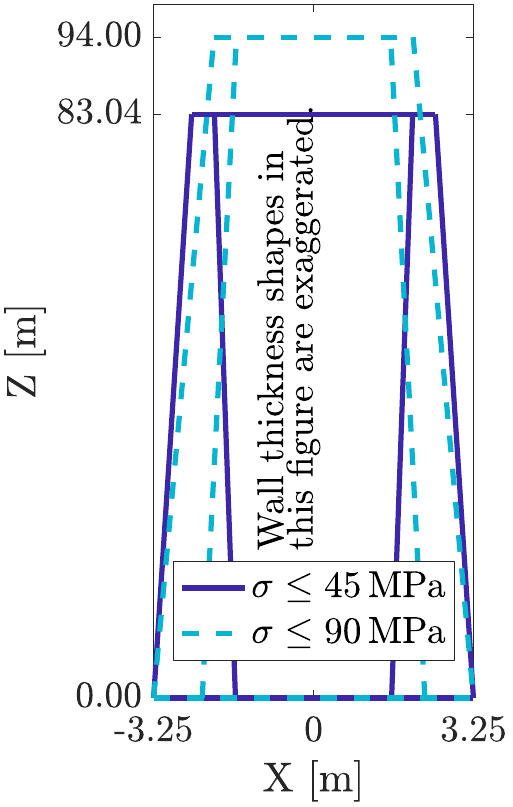}
    \caption{Optimal tower shapes for two distinct allowable tower stress values: 45 MPa and 90 MPa}
    \label{fig:result-tower-shapes_45_90}
\end{figure}

\begin{figure*}[ht!]
    \centering
    \subfloat{\includegraphics[width=5.15cm]{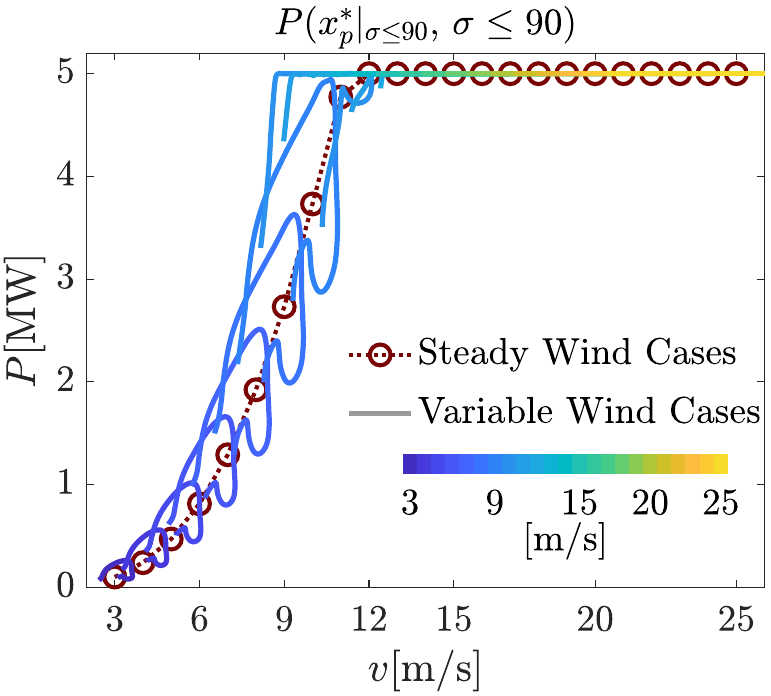}}
    \subfloat{\includegraphics[width=5.15cm]{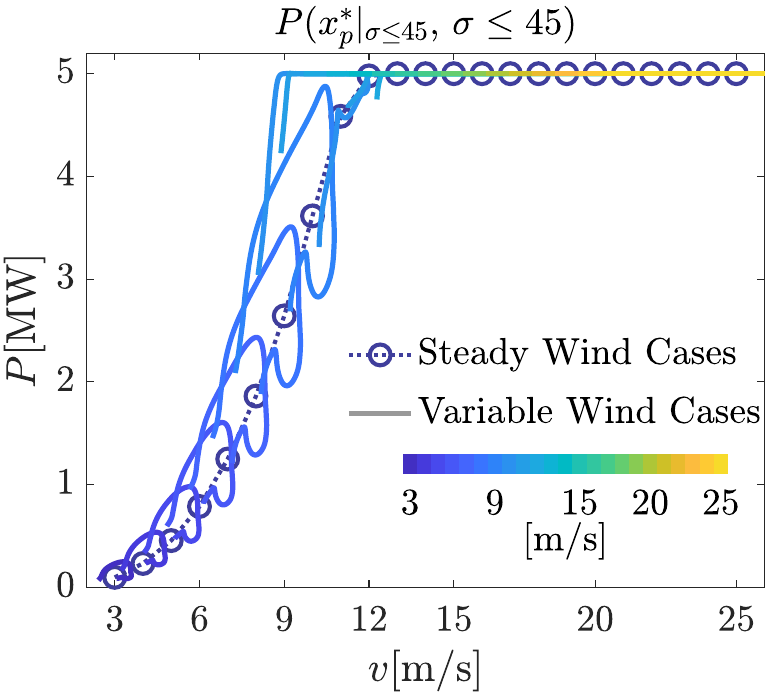}}
    \subfloat{\includegraphics[width=5.15cm]{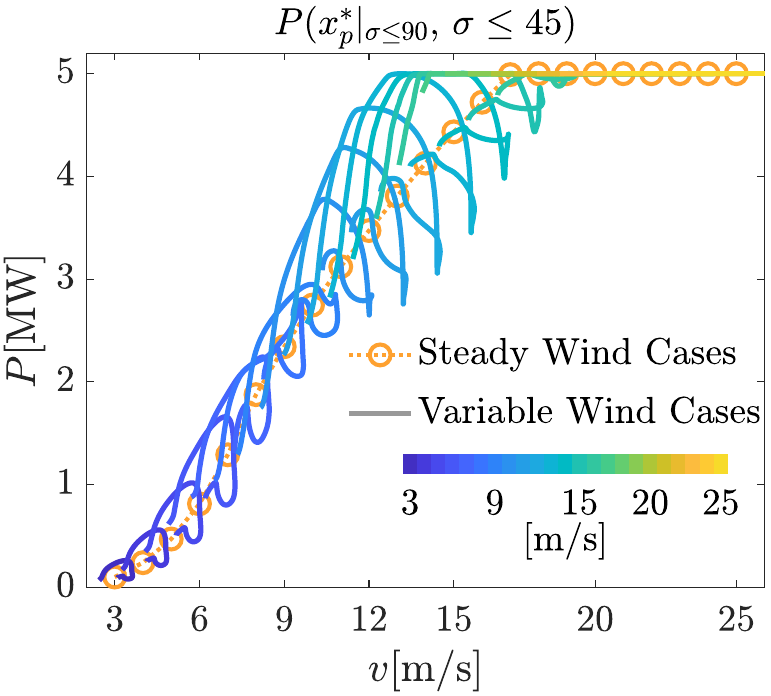}}
    \par
    \vspace{1em}
    \subfloat{\includegraphics[width=5.15cm]{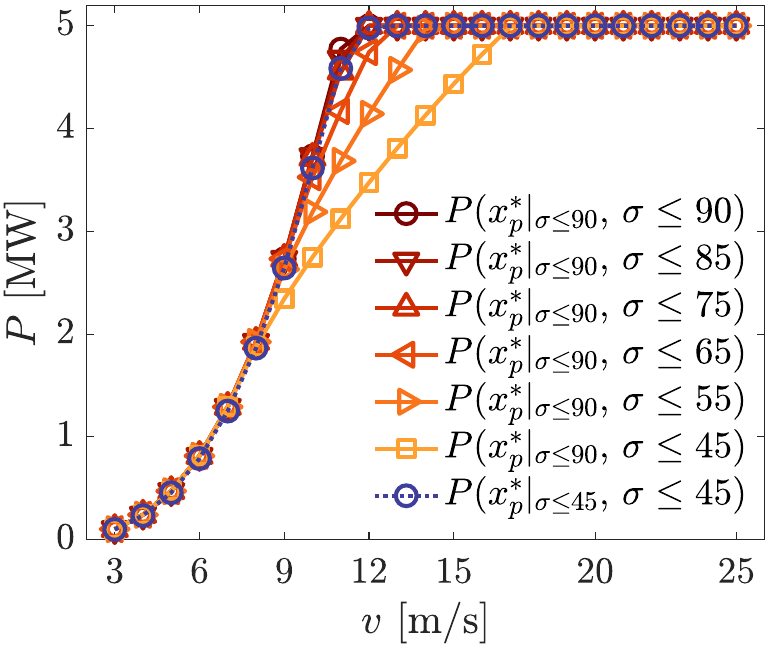}}
    \subfloat{\includegraphics[width=5.15cm]{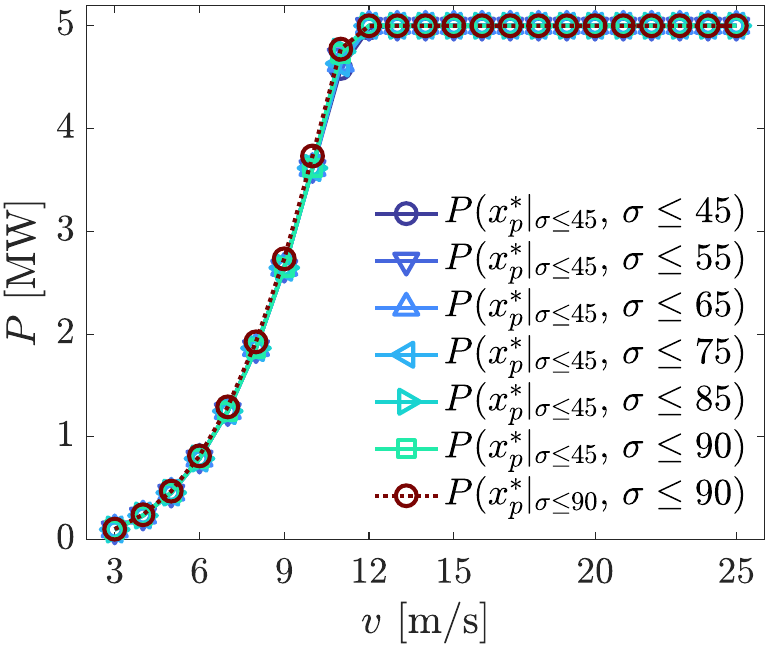}}
    \subfloat{\includegraphics[width=5.15cm]{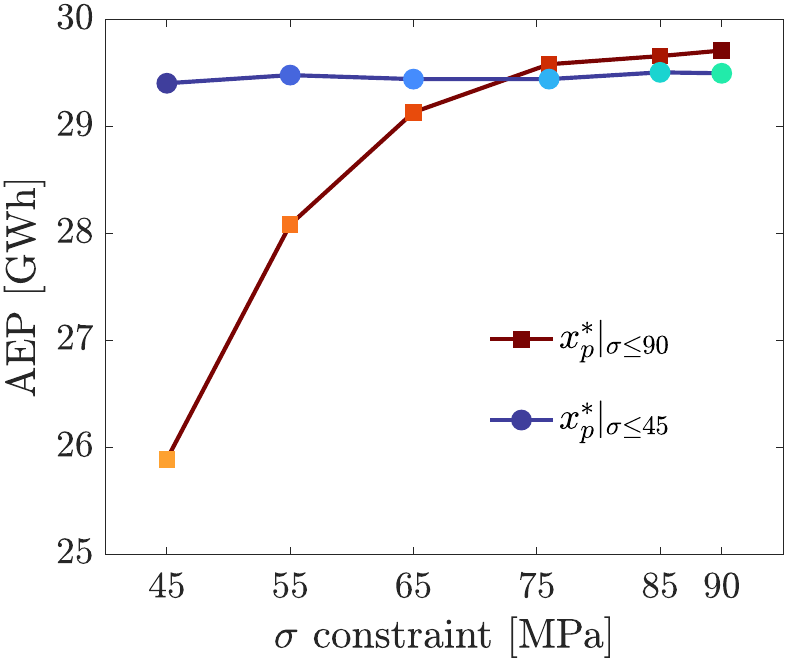}}
    
    \caption{Generated power curves under various scenarios. Sub-figures refer to power curves of steady and varied wind cases for (a)~$x_p^{\ast} |_{\sigma \le 90},\, \sigma \le 90$; (b)~$x_p^{\ast} |_{\sigma \le 45},\, \sigma \le 45$; (c)~$x_p^{\ast} |_{\sigma \le 90},\, \sigma \le 45$; Comparison of steady wind power curves with varied tower stress constraints in the control problem, when optimal plant design is obtained at (d)~$\sigma \le 90$ MPa; (e)~$\sigma \le 45$ MPa; and (f)~Comparison of AEP trends for cases illustrated in (d) and (e).}
    \label{fig:power-curve}
\end{figure*}

To quantify the effect of failure mode constraints on the CCD result, a new study is presented in Figs.~\ref{fig:result-tower-shapes_45_90}--\ref{fig:power-curve}. In this study, the maximum allowable tower stress is increased from 45 MPa to 90 MPa. This corresponds to a different (more expensive) tower material selection. Figure~\ref{fig:result-tower-shapes_45_90} shows the optimal tower shapes obtained for these two cases. Increased allowable stress results in optimal designs with increased tower heights, increased wind load, and a higher AEP. 

The study involves comparing power curves generated using two different stress upper bounds. Additionally, examining the changes in the power curve when the plant design is obtained based on one case, but the OLOC is solved assuming a maximum stress constraint from the other case would provide insightful results. This concept is common in many engineering situations, where plants are designed either using CCD or sequential approaches with certain constraints, but the maximum constraint may change in reality. This study demonstrates that such changes can degrade system response and performance. In this context, ${x^{\ast}_p}|_{\sigma \le \sigma_i}$ represents the optimal plant obtained using the CCD approach with a maximum stress constraint of $\sigma_i$ MPa. Subsequently, $\sigma \le \sigma_j$ indicates that the power curve is generated using the optimal plant from the previous step, while the maximum stress in the OLOC simulation is set to $\sigma_j$ MPa.

Figure~\ref{fig:power-curve}(a) illustrates the power curve obtained when the maximum tower stress is set to 90 MPa. In this case, ${x^{\ast}_p}|_{\sigma=90}$ refers to the optimal plant obtained using the CCD approach with a maximum stress constraint of 90 MPa. Subsequently, $\sigma=90$ indicates that the power curve is generated using the optimal plant from the previous step, while the maximum stress in the OLOC is also set to 90 MPa. In part (a) of the figure, both the CCD and OLOC utilize the same maximum stress constraint to generate the power curve, indicating an expected favorable outcome. The power curve is generated using a constant wind speed ranging from 3 to 25 m/s in 1 m/s increments. The simulation is conducted over a duration of 10 minutes, with the average power value computed. Additionally, to examine the power time series for non-constant wind speeds, the wind profile depicted in Fig.~\ref{fig:wind-function} is employed, incorporating average values between 3 and 25 m/s. As depicted in Fig.~\ref{fig:power-curve}(a), the power curve exhibits an increase roughly proportional to the cube of the wind speed in region 2 (below rated wind speed). In region 3 (above rated wind speed), the power remains relatively constant (assuming power constancy instead of torque constancy). Moreover, the power time series aligns closely with the average values derived from the constant wind speeds.

Figure~\ref{fig:power-curve}(b) depicts a similar plot, but in this case, both the CCD and OLOC simulations employ a constant stress of 45 MPa. Figure~\ref{fig:power-curve}(c) illustrates the results when the plant is obtained from the CCD case with a maximum stress of 90 MPa, but in the subsequent OLOC simulation, the stress constraint is reduced to 45 MPa. As observed, the power curves show a decline, which is reasonable considering the disparity between the maximum stress assumptions in the CCD and OLOC simulations.

In Fig.~\ref{fig:power-curve} (d), the power curve is presented for various cases. For the first six cases, the plant design is obtained from ${x^{\ast}_p}|_{\sigma=90}$, but the maximum stress in the OLOC simulation is varied from 45 to 90 MPa. It can be observed that decreasing the maximum stress in the OLOC simulations causes a downward shift in the power curve. Furthermore, as the maximum stress is reduced further, the power decrease becomes more pronounced. The last case in Fig.~\ref{fig:power-curve} (d) corresponds to ${x^{\ast}_p}|_{\sigma=45}$, and interestingly, the power curve for this case is nearly identical to that of ${x^{\ast}_p}|_{\sigma=90}$. The only distinction between these two cases lies in the assumption regarding the maximum stress in the CCD and OLOC simulations. In the last case, the maximum stress remains the same in both simulations, while in the previous case, the plant obtained from ${x^{\ast}_p}|_{\sigma=90}$ is designed to tolerate stresses ranging from 45 to 90 MPa. However, enforcing a maximum stress of 45 MPa in the OLOC simulation leads to performance degradation, requiring a compromise in power generation to meet the stress constraints.

Figure~\ref{fig:power-curve} (e)  illustrates a similar concept, but this time the plant design is obtained from ${x^{\ast}_p}|_{\sigma=45}$ for the first six cases. As observed, increasing the maximum stress in the OLOC simulation does not alter the power curve because the plant design is already based on an assumption of a maximum stress of 45 MPa. Therefore, increasing this constraint in the OLOC simulation has no effect on the power curve.

Figure~\ref{fig:power-curve} (f) displays the Annual Energy Production (AEP) derived from parts (d) and (e) of the analysis. For ${x^{\ast}_p}|_{\sigma=90}$, reducing the maximum stress constraint results in a degradation of the power curve and consequently a decrease in AEP. However, for ${x^{\ast}_p}|_{\sigma=45}$, increasing the maximum stress bound does not alter the power curve significantly, and the AEP remains almost the same. 
Upon examining this figure, one may question the purpose of employing a higher maximum stress constraint in the CCD if the power curve obtained from a 45 MPa assumption is not considerably lower than the 90 MPa case. The response to this query is two-fold:
\begin{enumerate}
    \item As illustrated in Fig.~\ref{fig:result-tower-shapes} the tower designed for ${x^{\ast}_p}|_{\sigma=45}$ is thicker, potentially increasing the tower mass and subsequently the total cost. In this study, for instance, the tower mass increased from 244.9 tonnes in ${x^{\ast}_p}|_{\sigma=90}$ to 344.4 tonnes in ${x^{\ast}_p}|_{\sigma=45}$.
    \item Even a small enhancement in AEP can yield significant monetary benefits. In the present case, the difference amounts to approximately 0.5 GWh. Assuming an electricity cost of $\$0.13$ per kWh, this translates to an annual savings of approximately $\$65,000$ USD.
\end{enumerate}
Overall, the consideration of factors such as tower mass and cost, as well as the potential financial gains from even slight improvements in AEP, justifies the exploration of higher maximum stress constraints in the CCD optimization process.

\subsection{Comparison Studies Showing Tower Mass Effect on AEP}
\label{Comparison Studies Showing Tower Mass Effect on AEP}

\begin{table}[ht!]
\centering
\caption{Comparisons of parameters and results for OLOC studies with optimal tower mass (Case~1), increased tower mass (Case~2), and forward ordinary differential equation (ODE) simulation result using the control trajectories of Case~1 with the tower design of Case~2 (Case~3). $P_{\text{out}}$ is reported for the case where wind speed is equal to 14 m/s.}
\label{tab:comparison}
\begin{tabular}{cccc}
\toprule
Parameters& Case 1 & Case 2 & Case 3 \\
\midrule
$t_{\text{base}} \text{ [m]}$ & 0.042 & 0.042  & 0.042\\
$t_{\text{tip}} \text{ [m]}$ & 0.012 & 0.028  & 0.028\\
$d_{\text{tip}} \text{ [m]}$ & 4.95 & 6.50  & 6.50\\
$l \text{ [m]}$ & 83.04 & 83.04  & 83.04\\
$m_{\text{t}} \text{ [kton]}$ & 0.345 & 0.505  & 0.505\\
$m_{\text{p}} \text{ [kton]}$ & 7.472 & 7.311  & 7.311\\
$P_{\mathrm{out}} \text{[MW]}$ & 4.93 & 4.83  &  4.83\\
$\text{AEP} \text{ [GWh]}$ & 29.22 & 28.73 & \\
Method & OLOC & OLOC & Simulation \\
\bottomrule
\end{tabular}
\end{table}

\begin{figure}[ht!]
    \centering
    \includegraphics[width=1.0\linewidth]{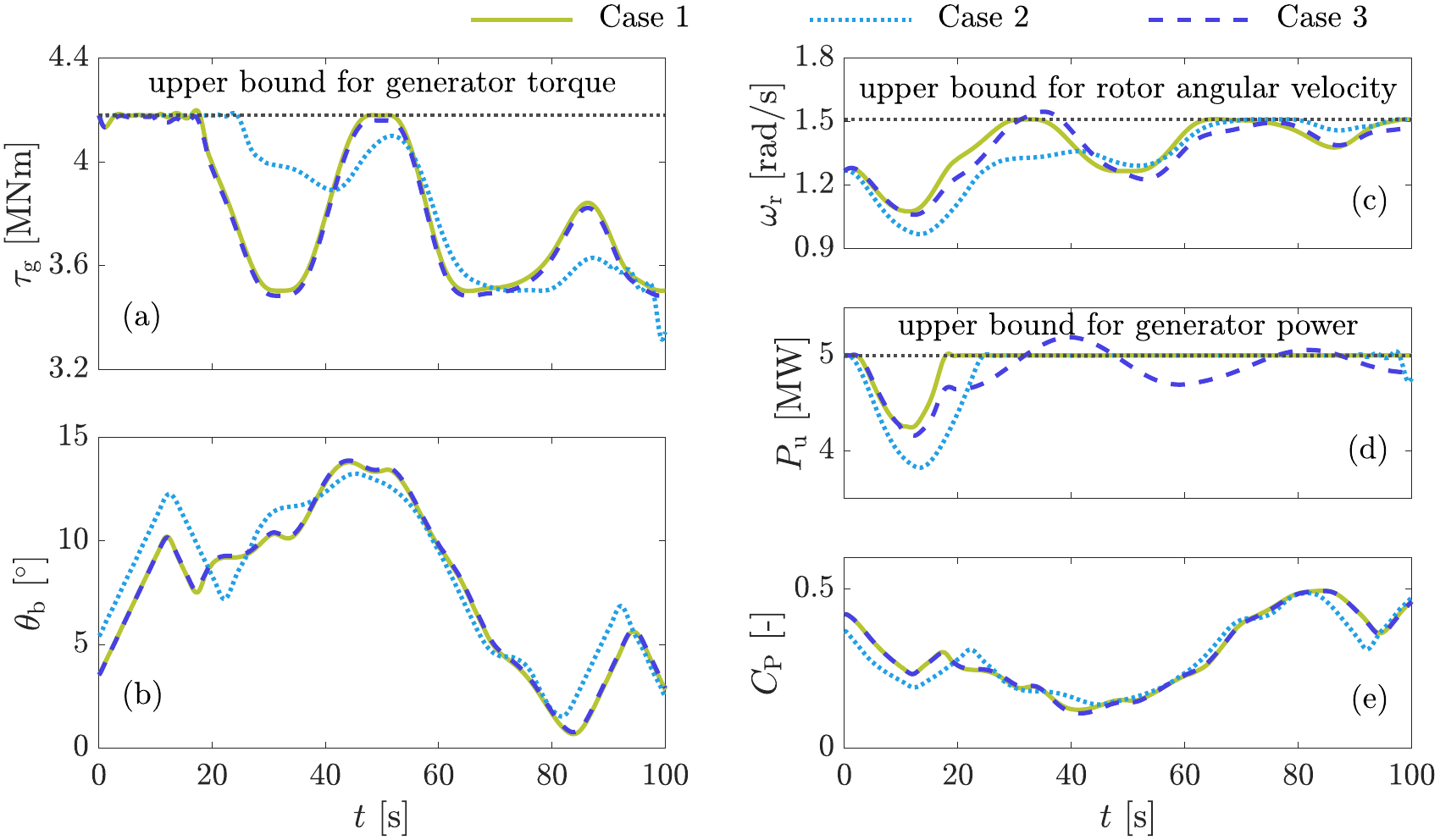}
    \caption{Trajectories of control inputs and model outputs that correspond to the three cases listed in Table~\ref{tab:comparison}. Dotted horizontal lines indicate the upper bounds for corresponding quantities. (a)~Generator torque; (b)~blade pitch; (c)~Rotor rotational speed; (d)~Generator power; (e)~Power coefficient.}
    \label{fig:comparison-result-1-2}
\end{figure}


    

Another set of comparison studies was performed to show the effect of changing tower mass on AEP while maintaining a consistent tower height. Table~\ref{tab:comparison} contains the summary of parameter values and the results of this comparison study. Case~1 is the OLOC problem with the optimal tower design parameters obtained via the CCD problem presented above. Case~2 is another OLOC problem with the same tower height but with increased tower mass. Case~3 is based on solving the  dynamic system model (ordinary differential equations) using a forward simulation and the control trajectories obtained from Case~1 and the tower design of Case~2. All three problems are solved using the same wind profile with an average wind speed $\bar{u}$ of 14 m/s. In summary, Cases~1 and 2 are solved for obtaining the optimal control trajectory designs, while Case~3 is solved using a forward simulation with provided control trajectories.

\begin{figure}[ht!]
    \centering
    \includegraphics[width=1.0\linewidth]{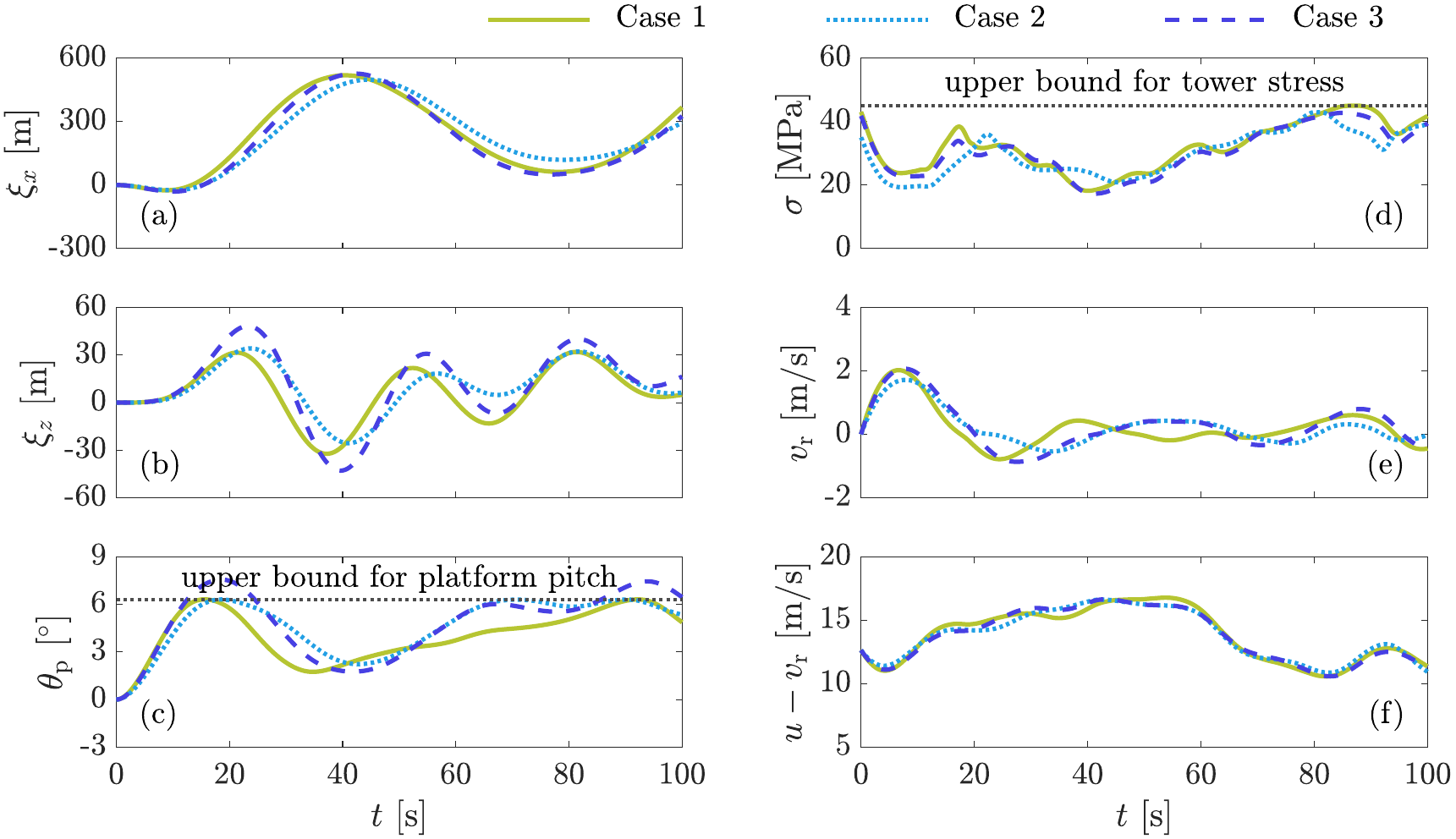}
    \caption{Trajectories of model dynamic outputs, including motion trajectories, stress, and velocities that correspond to the three cases listed in Table~\ref{tab:comparison}. Dotted horizontal lines indicate the upper bounds for corresponding quantities. (a)~Platform surge motion; (b)~platform heave motion; (c)~platform pitch motion; (d)~tower stress; (e)~rotor hub fore-aft velocity; (f) relative wind velocity.}
    \label{fig:comparison-result-3-4}
\end{figure}

Comparisons of results for these three cases are shown in Figs.~\ref{fig:comparison-result-1-2}--\ref{fig:comparison-result-3-4}. Figure~\ref{fig:comparison-result-1-2}(a) and (b) shows the generator torque and blade pitch trajectories, which are controlled quantities. Actual control signals used in numerical optimization are rates of these quantities. The control trajectories for Case~3 are extracted from the OLOC result of Case~1. However, due to a difference in numerical integration schemes, generator torque trajectories exhibit a slight difference. Case~1 uses the Gaussian quadrature method for the numerical integration, while Case~3 uses the \citet{Dormand1980ODE45} method, a family of higher-order Runge-Kutta methods. Notably, the optimal control trajectories for Cases~1 and 2 are significantly different since the tower designs for these cases are different. This difference indicates that the control strategy should be individually established for a specific plant design, which could be attainable using the CCD strategy presented in this study. Feedback control laws could be synthesized based on the OLOC CCD results. It is possible that for some CCD problems, no feedback control exists that can mimic the behavior of the OLOC trajectories. Methods for systematically informing CCD optimization studies based on feedback control law limitations is a topic of ongoing work. 

Figure~\ref{fig:comparison-result-1-2}(c)--(e) shows the trajectories of rotor rotational velocity, generator power, and power coefficient curves, and Fig.~\ref{fig:comparison-result-3-4} shows the trajectories of model dynamic responses, including platform motions in (a)--(c), tower stress in (d), and velocities in (e)--(f). The plant design for Case~1 is selected from the `Tower only' CCD result, which is already optimized and is given in Tab.~\ref{tab:optimized-parameters}. In contrast, the plant design for Case~2 deviates from the optimal design since tower mass was intentionally increased. Thus, in Fig.~\ref{fig:comparison-result-1-2}(d), the area under the generator power curve $P_{\text{u}}$ for Case~1 is significantly larger than for Case~2. Also, Case~1 exhibits a much longer period of constraint saturation (active path constraint indicated by the dotted horizontal line) compared to Case~2, as shown in Fig.~\ref{fig:comparison-result-1-2}(d). In addition, because the control trajectories used in Case~3 are not optimized with respect to its unique plant design parameters, we see that the Case~3 simulation result violates bounds and path constraints, as shown in Figs.~\ref{fig:comparison-result-1-2}(c)--(d) and \ref{fig:comparison-result-3-4}(c).

From the results given in Figs.~\ref{fig:comparison-result-1-2}--\ref{fig:comparison-result-3-4}, increased tower mass produces a larger tower mass moment of inertia with respect to the rotational center for the platform pitch DOF. Thus, Case~3 exhibits a larger value in the peak platform pitch motion, as shown in Fig.~\ref{fig:comparison-result-3-4}(c). As Cases~2 and 3 have common plant designs, the OLOC result from Case~2 can provide a mitigating strategy for the increased platform pitch motion from Case~3. By comparing the control trajectories of Cases~2 and 3, we observe that the blade pitch motion in Case~2 is significantly increased, resulting in reduced aerodynamic thrust applied to the turbine. With this optimized control strategy, the platform pitch motion can be mitigated to satisfy the constraint with the increased tower mass in Case~2, but with a sacrifice in power generation.

We used the same absolute wind profiles at the same mean wind velocity value ($\bar{u} = 14$ m/s) for fair comparisons across these three cases. One item to note is that the relative wind velocity profiles for these three cases are different from each other, as shown in Fig.~\ref{fig:comparison-result-3-4}(f). The reason behind this difference is that the thrust load exerted to the wind turbine is based on the relative wind velocity considering the rotor hub fore-aft motion. Thus, the relative wind velocity is calculated by subtracting the rotor hub fore-aft velocity from the absolute wind velocity. Due to significant platform pitch motion (unlike land-based wind turbines), the rotor hub fore-aft velocity is significant. As a result, the relative wind velocity profile can vary depending on plant design, and this variation impacts the resulting optimal control strategy and power generation.

It should also be noted that the platform shape across all studies presented here is fixed. The only variable here that can change the platform properties is the variable ballast mass. When the tower design parameters are modified, the variable ballast mass is automatically modified to satisfy the mass and buoyancy equilibrium requirement in the platform heave DOF. However, there are a few other ways to achieve mass and buoyancy equilibrium: (1) adjust the submerged volume of the platform by changing shape, or (2) adjust equilibrium draught height. Each of these options may affect optimization results significantly. However, to maintain simplicity, we kept the platform shape and the equilibrium draught height as fixed constants. Instead, we defined the variable ballast mass as a variable that depends on the tower mass. More flexible design studies may reveal new insights and potentially higher performance. 

This study highlights an important finding that increasing tower mass does not necessarily result in an increase in generated power and Annual Energy Production (AEP). In fact, it was observed that for a fixed tower length, increasing the tower diameter and tip thickness led to a worse outcome in terms of power generation. This underscores the need for an integrated approach in obtaining optimal plant design variables, considering the coupling between the design variables and the controller. By taking full advantage of this coupling, it becomes possible to optimize both the structural aspects and the controller, leading to improved overall performance and energy production.

\subsection{Wave and Fatigue Study}
\label{sec:Wave_and_Fatigue_Study}

\begin{figure*}[ht!]
    \centering
    \includegraphics[width=0.95\textwidth]{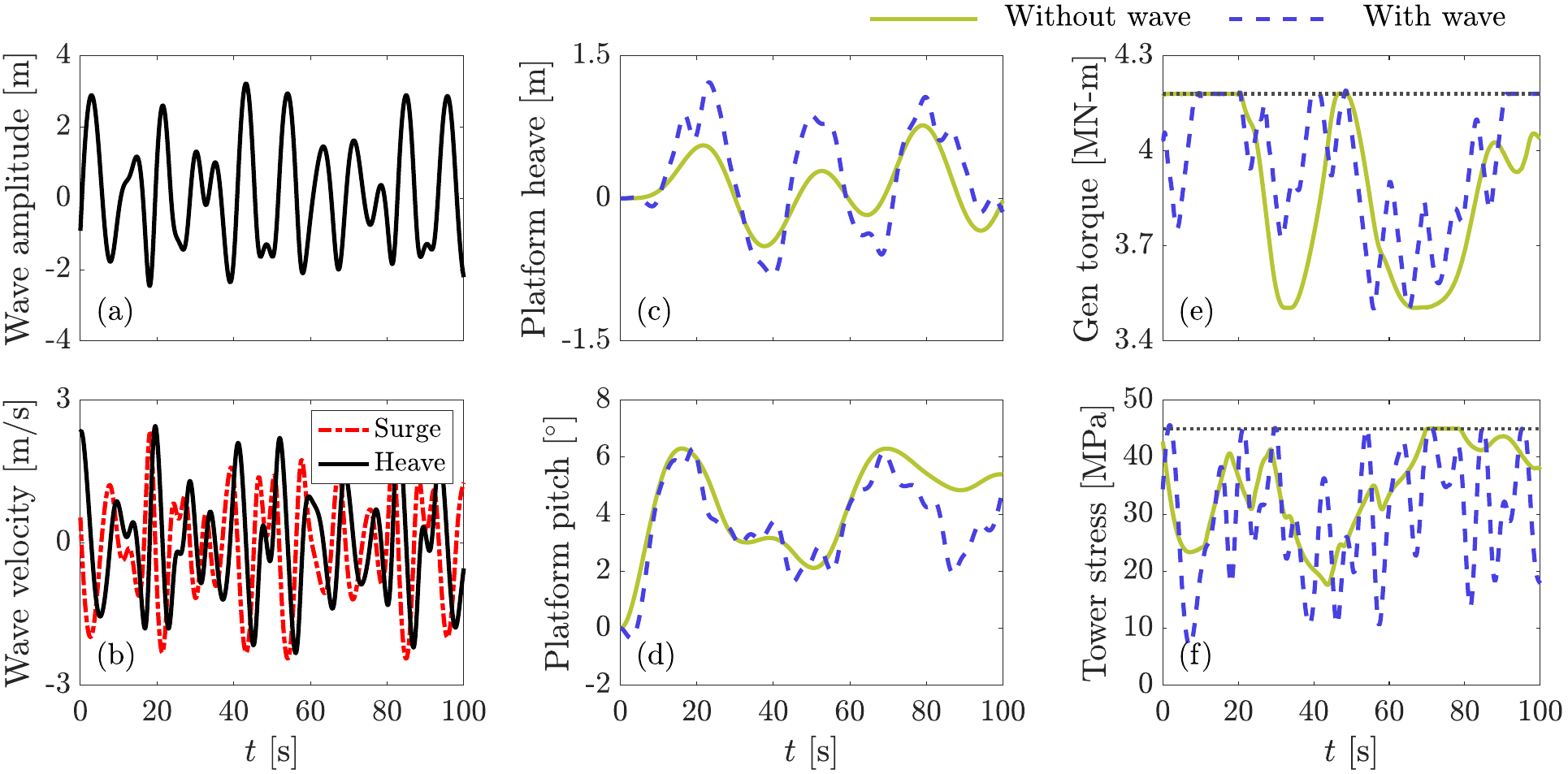}
    \caption{Optimal control performed on the ``tower only'' plant design given in Tab.~\ref{tab:optimized-parameters} without and with additional wave loading. (a) Wave amplitudes; (b) Wave velocities in surge and heave directions; (c) Platform heave motions; (d) Platform pitch motions; (e) Optimal generator torque trajectories; (f) Tower stress trajectories.}
    \label{fig:wave_study}
\end{figure*}

In the previous sections, the dynamics of the FOWT system took into account the interaction between the water and the platform motion, but wave loads were excluded for simplicity. However, wave loading is an essential external disturbance that can cause FOWT system failures.
Wave loading leads an additional computational burden in solving the CCD problem of the FOWT system. Thus, instead of solving all CCD cases with wave loading, an optimal result obtained for the tower-only case with wave loading is compared to the result of the same problem without wave loading obtained in the previous section. Fatigue analysis is also performed on the tower-only CCD cases (with and without wave loading).

Figure~\ref{fig:wave_study} shows the optimal control results for two cases: without and with wave loading. The wave is obtained from \citet{al2017dynamics} and is a irregular signal with a period ($T_p$) of $10$ s and significant wave height ($H_s$) of $6$ m. Including wave loading results in higher frequency motion responses, requiring higher frequency control inputs. This may accelerate fatigue damage. To see this effect, Mlife \citep{hayman2012mlife}, a fatigue prediction code, is used to compute fatigue life of the tower. The lifetime is considered 20 years, and the ultimate strength is set in such a way that 20 years life time is satisfied. For the case that wave loading is not included, 20 years of lifespan is satisfied with the tower design strength of $45.08$ MPa, which is close to the previous CCD result without considering fatigue lifespan ($45$ MPa). However, when the wave loading is implemented, the required tower design strength needs to be increased to $61.78$ MPa to satisfy the 20 years of lifespan. It should be noted that this is an initial study utilizing low-order models to decrease computation time. More detailed models need to be incorporated to verify fatigue lifespan of the entire FOWT system design.

\section{Conclusion}
\label{sec:conclusion}

In this article, a nested CCD method using OLOC is implemented to solve an integrated FOWT design optimization problem. The reduced order FOWT dynamic model is based on a spar buoy floating platform and the NREL 5-MW reference turbine. Tower and blade designs are parameterized using 14 plant design variables for the outer-loop problem. The time-domain OLOC problem is formulated using the pseudospectral method for the inner-loop problem. The objective function for this nested CCD problem is to maximize the AEP, subject to constraints defined to enforce stress, motion, and other various operating limits. Neural network-based mooring line dynamic models are trained for rested and suspended mooring line conditions to facilitate efficient computation for use with optimization studies. The simplified reduced-order modeling approach enabled efficient CCD studies that are practical to solve using standard computational resources. At the same time, the model fully captures all of the essential physics behind the dynamic behaviors of the FOWT, especially those that give rise to key design coupling relationships.

The CCD result shows an increase in the AEP compared to the baseline design by more than eleven (11) percent. In addition, we demonstrated a possibility to expand this CCD problem with various additional objective functions, including tower mass, to form a multiobjective optimization problem to help reduce the cost of the FOWT system further. Influences of the design variables on the objective function are investigated using sensitivity analysis. By comparing the inner-loop OLOC solutions for increased tower mass design, we identified a dynamic mechanism that leads to a certain optimal tower mass that produces maximal power. Due to platform pitch motion, there exists a sweet spot for the optimal tower mass and height that maintains structural integrity and limits platform motion while supporting large wind speeds at high rotor hub elevations.

Furthermore, as explored in Section~\ref{sec:omparison of Increased Maximum Allowable Tower Stress}, it was demonstrated that the power curve and AEP are influenced by the maximum tower stress constraint used in CCD. It was observed that if the maximum stress in the OLOC simulation is lower than the assumption made in CCD, it results in a degradation of the power curve and AEP. Conversely, if the maximum stress in the simulation is higher than the assumption made in CCD, the power curve remains relatively unchanged, but it leads to an increase in tower mass and consequently higher capital costs. Additionally, even when the power curves are closely aligned, even a small difference between them can translate into substantial financial discrepancies. Therefore, these types of constraints in CCD should be set in a manner that accurately represents real-world conditions, with an added margin of safety.

The nested CCD implementation presented here (based on a simplified reduced-order FOWT dynamic model) is appropriate for early-stage FOWT design studies in practice. CCD studies based on this implementation can reveal overall preferred design directions for system designers to explore. Also, this implementation can be used as a template for developing more advanced CCD approaches that incorporate higher fidelity FOWT system models, such as OpenFAST \citep{OpenFAST} simulation code, as well as enhanced design flexibility/representation fidelity. An additional ongoing effort to improve the practicality of similar CCD implementations is to adapt closed-loop control (CLC) or MPC design capabilities in the inner-loop optimal control problem. In this way, the ideal OLOC design results can inform promising directions for engineers to explore for more practical CLC or MPC-based controllers.

\section*{Funding}

The information, data, or work presented herein was funded in part by the Advanced Research Projects Agency-Energy (ARPA-E), U.S. Department of Energy, under a project titled: ``WEIS: A Tool Set to Enable Controls Co-Design of Floating Offshore Wind Energy Systems'' associated with award no DE-AC36-08GO28308. 

\section*{Declaration of Competing Interest}

The authors declare that they have no competing interest.

\section*{Data Availability}

All data required to replicate the results can be generated by the MATLAB optimization code. A MATLAB optimization code for a select problem demonstrated in the manuscript are available upon request to the first or corresponding authors. Readers may need a commercial pseudospectral optimal control solver, GPOPS-II, to run the provided MATLAB codes.

\bibliographystyle{elsarticle-num-names} 
\bibliography{main}
\end{document}